\documentclass[aps,showpacs,longbibliography,notitlepage,superscriptaddress,twocolumn,nofootinbib]{revtex4-1}
\pdfoutput=1
\usepackage[utf8]{inputenc}
\usepackage[english]{babel}
\usepackage[T1]{fontenc}
\usepackage{amsmath}
\usepackage{xcolor}
\colorlet{myPurple}{blue!40!red}
\colorlet{myCyan}{cyan!60!gray}
\colorlet{myRed}{blue!55!gray}
\usepackage{tikz}
\usepackage{pgfplots}
\pgfplotsset{compat=1.14}
\usepackage[colorlinks=true,citecolor=myRed,urlcolor=myRed,linkcolor=myRed]{hyperref}
\usepackage{exscale}
\usepackage{fouriernc}
\usepackage{bbm}
\usepackage{graphicx}
\usepackage{amsmath}
\usepackage{latexsym}
\usepackage{amssymb}
\usepackage{mathtools}
\usepackage{amsfonts}
\usepackage{physics}
\usepackage{amssymb}
\usepackage{times}
\usepackage[T1]{fontenc}
\usepackage{amsthm}
\usepackage{enumerate}
\usepackage{csquotes}
\usepackage{bbold}
\usepackage{color}
\usepackage{nicefrac}
\usepackage{changes}
\usepackage{calrsfs}
\usepackage{soul}

\newcommand{\avg}[1]{\langle{#1}\rangle}

\usepackage{tcolorbox}

\newcommand{\argmin}{\operatornamewithlimits{argmin}}

\usepackage{tikz}
\usepackage{lipsum}
\theoremstyle{plain}

\usepackage{graphicx}
\usepackage{bm}
\usepackage{dsfont}
\usepackage{tikz}
\usepackage[T1]{fontenc}
\usepackage{amsthm}
\usepackage{array}
\usepackage{amssymb}
\usepackage{amsfonts}
\usepackage{cancel}
\usepackage[toc,page]{appendix}
\usepackage{multirow}
\usepackage{color}
\usepackage{calrsfs}
\usetikzlibrary{backgrounds,decorations.pathreplacing,calc}
\definecolor{mygray}{gray}{0.8}
\DeclareMathAlphabet{\mathcal}{OMS}{cmsy}{m}{n}

\begin{document}
\title{Quantum verification and estimation with few copies}

\author{Joshua Morris}`
\affiliation{University of Vienna, Faculty of Physics, Vienna Center for Quantum Science and Technology (VCQ), 1090 Vienna, Austria}
\author{Valeria Saggio}
\affiliation{University of Vienna, Faculty of Physics, Vienna Center for Quantum Science and Technology (VCQ), 1090 Vienna, Austria}
\author{Aleksandra Go\v{c}anin}
\affiliation{Faculty of Physics, University of Belgrade, Studentski Trg 12-16, 11000 Belgrade, Serbia}
\author{Borivoje Daki\'{c}}
\affiliation{University of Vienna, Faculty of Physics, Vienna Center for Quantum Science and Technology (VCQ), 1090 Vienna, Austria}
\affiliation{Institute for Quantum Optics and Quantum Information (IQOQI),
Austrian Academy of Sciences, Boltzmanngasse 3, 1090 Vienna, Austria}

%\date{\today}

\begin{abstract}
As quantum technologies advance, the ability to generate increasingly large quantum states has experienced rapid development. In this context, the verification and estimation of large entangled systems represents one of the main challenges in the employment of such systems for reliable quantum information processing. Though the most complete technique is undoubtedly full tomography, the inherent exponential increase of experimental and post-processing resources with system size makes this approach infeasible even at moderate scales. For this reason, there is currently an urgent need to develop novel methods that surpass these limitations. This review article presents novel techniques focusing on a fixed number of resources (sampling complexity), and thus prove suitable for systems of arbitrary dimension. Specifically, a probabilistic framework requiring at best only a single copy for entanglement detection is reviewed, together with the concept of selective quantum state tomography, which enables the estimation of arbitrary elements of an unknown state with a number of copies that is low and independent of the system’s size. These hyper-efficient techniques define a dimensional demarcation for partial tomography and open a path for novel applications.
\end{abstract}

\maketitle

\section{Introduction}
\label{introduction}

In the coming decades, thanks to rapid technological advances, the probability of a new information revolution appears quite high. Quantum systems involving photons, atoms, spins, molecules, solid-state and optomechanical devices, even with the absence of perfect control and manipulation, are already promising candidates for building new applications aside from universal quantum computing. As difficult as it is to predict how emerging technologies will be most effectively applied, one can expect to see quantum technologies with a high degree of variability in architecture and capacity (as when classical computers emerged in the 1950s), the so-called noisy, intermediate-scale quantum (NISQ)~\cite{preskill2018quantum}. Here intermediate-scale refers to the size of the quantum processors, in the regime of tens of qubits up to a few hundred in the next decade or so. Remarkable achievements in creating larger quantum states have already been reported~\cite{kelly2019operating,friis2018observation,zhong2020quantum,wang2019boson, gong2021quantum, ebadi2021quantum, mooney2021wholedevice} using different quantum platforms, from superconducting architectures to trapped ion systems and photonic setups. 
Moreover, impressive demonstrations (such as those of a computational quantum advantage) have recently been reported by several groups that used 53~\cite{arute2019quantum} and 56~\cite{wu2021strong} superconducting qubits and up to 113 photons~\cite{zhong2020quantum,zhong2021phase}.

Such rapid development and demonstration of a quantum supremacy indicate that quantum information processing is sufficiently mature that another problem, quite aside from noisy quantum systems, has begun to make its presence known with increasing frequency. While it is all very well to coherently process quantum states that reside in an exponentially large space, it means little if one cannot retrieve and validate the results of such manipulations. So begins consideration for the metrology of quantum systems. The gold standard of quantum measurement is full state tomography \cite{james2005measurement}, wherein complete knowledge about the state is gained via measurement. Though certainly sufficient, the complexity in both measurements and computational processing power grows exponentially fast with the dimension of the quantum system. 

Given that our interest in quantum information processing is this rapid growth, inserting a step that requires exponential resources seems rather counterproductive. Until very recently, however, this exponential cost was largely irrelevant as our ability to rapidly measure or classically compute vastly outstripped our ability to perform meaningful operations with more than a few qubits. Thus, simply performing full state tomography and retrieving a complete quantum state was a viable strategy. This approach was only ever practical at the very small scale of NISQ and pre-NISQ however. In the long term, fault-tolerant and noise-resistant quantum computers ought to make a complete validation of the system less important but we are far away from such feats of quantum engineering, while still being capable of constructing large quantum systems. Thus a gap has appeared - systems are too large for anything nearing complete tomography but not advanced enough to assume low errors.

The advantages of a complete tomography are obvious. One need make no assumptions on any properties of the target system except that it can be repeatedly produced (reinitialised) and measured. The price of such ignorance is an exponential cost in measuring, reconstructing and storing the state of the target and is naturally unsustainable as we move into the intermediate regime. But such a problem has hardly taken the quantum estimation community by surprise and many strategies exist to mitigate such a heinous complexity cost. 
Often, complete information is not required in many cases and when married with random sampling techniques can result in powerful verification methods \cite{knill2008randomized, gross2010quantum,pappa2012multipartite,tran2015quantum,montanaro2017learning, pallister2018optimal, torlai2018neural, zhu2019efficient, kueng2018,kueng2019,dimic2018single, saggio2019experimental, huang2020predicting, zhu2019optimal, dimicsupicdakic, aaronson2017, morris2019selective} (see also~\cite{Eisert2020} for general review on the topic) that probe only some specific quantities one might wish to know about a given state. To name but a few, one might wish to investigate only the presence of entanglement in a certain quantum state~\cite{dimic2018single, saggio2019experimental} or directly estimate the state fidelity~\cite{flammia2011direct}, i.e., the quantification of the overlap between prepared and ideal states. It follows naturally that reducing the amount of obtainable information comes with a lower demand in terms of experimental resources, thus making these methods more viable alternatives when the full density matrix is not needed. For clarity, we will explicitly define here that any interrogation of a quantum system which reveals information about that system is termed a partial tomography. 

It appears then that a trade-off of some kind must occur. Complexity costs can be reduced in one regard but increased in another~\cite{Eisert2020}, essentially shifting the difficulty to another stage of an experiment, or we can reduce the information extracted. Ultimately, an explicit dimension dependence remains in most tasks and this serves as a problematic complication for large-scale systems. With this in mind, we concern ourselves with strategies that appear to saturate some notion of maximal information extraction, paired with a resource cost (at every stage of the protocol) that has moderate growth in the dimension. This suggests a different mode of thinking may be in order. Rather than asking how large a quantum system we can effectively probe with a given strategy, consider instead the central question of this review: 

\begin{displayquote}
\textit{Given a limited number of interactions with a large system, how much classical information can we learn with a high degree of certainty?}
\end{displayquote}

This extracted classical information can take many forms and one must be careful of the kinds of questions one asks. Consider the task of entanglement detection, which may be performed indirectly by estimating the mean value $\expval{W}$ of an appropriate witness $W$ and comparing it to some threshold value $W_c$, which requires repeated measurements on large ensembles of identically prepared quantum states. An alternative to this is a direct approach by an oracular question ``Is  $\expval{W}<W_c$?'', which potentially can be queried with a single copy. For detecting entanglement they of course produce the same answer, but estimating the expectation value is far more resource-intensive than bounding it from above in the first place. The benefit of doing so is clear, however, the question then is how to operationally reformulate the former into the latter. This process of reformulation is one of the central topics that shall form this review.

Such thinking engenders a curious divergence from the norm of quantum metrology wherein both the dimension of a system and the number of copies are seen as a given and large. On the other hand, this decision-theory centric approach, that has estimation as comparable to traversing a finite tree of outcomes to arrive at a final conclusion has been shown \cite{kueng2019, dimic2018single, saggio2019experimental, huang2020predicting, zhu2019optimal, dimicsupicdakic, aaronson2017, morris2019selective}
to yield vastly improved complexity bounds for previously challenging measurement tasks. 

\begin{figure*}[ht!]
\centering
\includegraphics[width=0.7\textwidth]{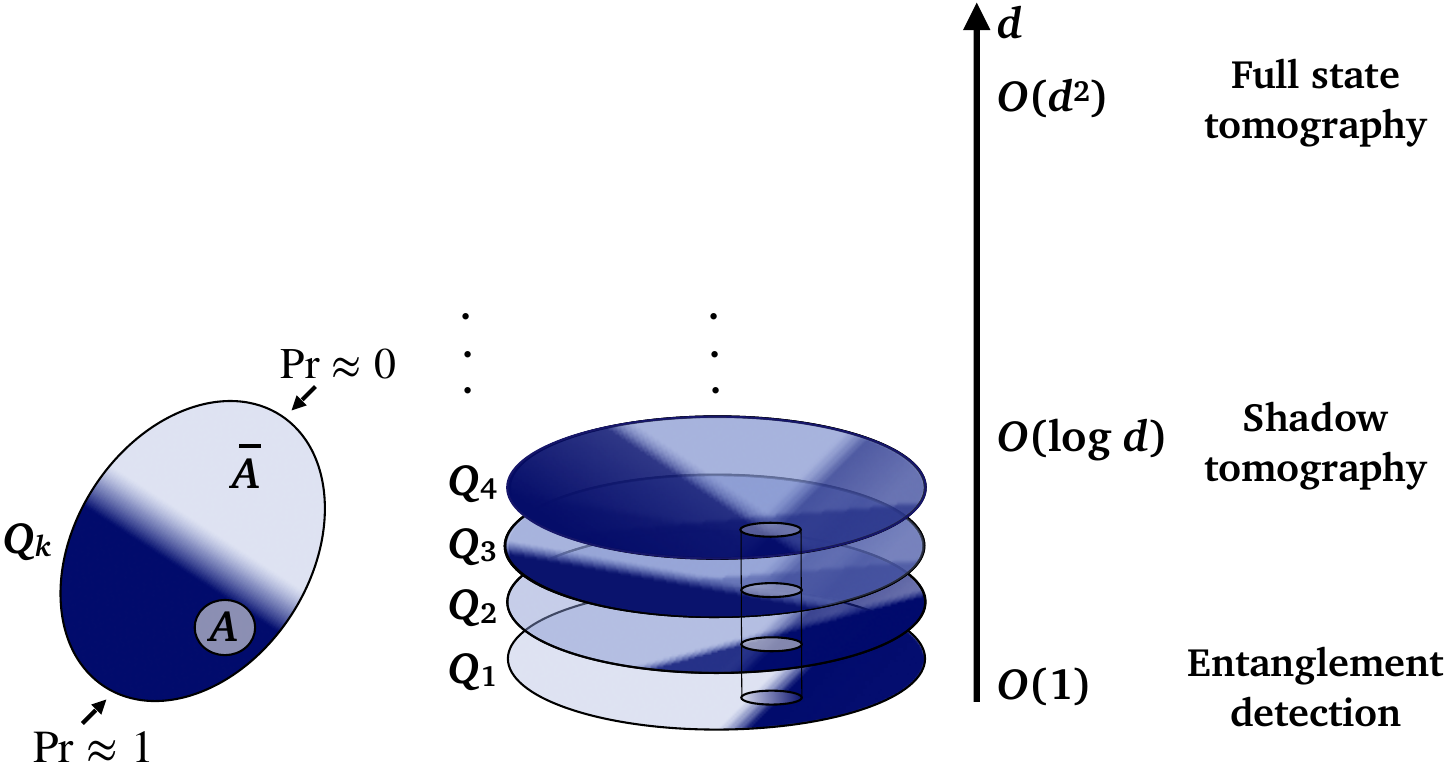}
\caption{\protect\raggedright \textbf{Schematic of the probabilistic procedure.} The probability $Pr$ that the quantum system contains the property $A$ is found by asking relevant questions $Q_k$ to the system. A probability close to $1$ is indicated by a dark region, in contrast to a probability close to $0$, associated to a lighter colour. Asking more and more questions builds up the probability that the system contains $A$.}    
\label{binary_fig1}
\end{figure*}

By rephrasing the problem of verification in this decision-theoretic way we define our starting condition as the resources of an efficient strategy, such as a limited number of state copies, and then list measurement protocols that operate within these constraints. As an illustration of the method, consider testing some property with $N$ copies available, where $N$ is potentially low (e.g., few copies). Each copy may then be considered as a precious resource for measurements we are permitted to ask a quantum system in order to ascertain its properties. For example, we wish to test if the state $\rho\in A$ or $\rho\in\bar{A}$ (with $A\cup \bar{A}$ being the complete set of states) where $A$ denotes the property being tested (as in Figure \ref{binary_fig1}). An efficient strategy is one where the queried system is overwhelmingly unlikely to pass a test condition if it does not contain the queried property $A$.

 The strategy is as follows. A set of carefully designed and easy-to-perform measurements $\mathbf{Q}=\{Q_1, Q_2, Q_3 \dots Q_L\}$ that serve as queries to the system are constructed. For the $k$th instance of the $N$ copies of a state, a query $q_k\in\mathbf{Q}$ is randomly chosen and applied to that instance, producing a sequence of query outcomes $\mathbf{i}=(i_1,...,i_N)$ for $i_k \in \{0,1\}$. This sequence together with the sequence of chosen queries $\mathbf{q}=(q_1,...,q_N)$ is then passed to a decision (cost) function $S(\mathbf{q},\mathbf{i})$ which produces a pass/fail result. We define a strategy to be efficient if it satisfies the following probabilistic expression
\begin{equation}\label{eff_inequality}
\Pr\left[S(\mathbf{q},\mathbf{i})=\mathrm{"pass"}|\rho \in \bar{A}\right] \leq \exp[-\alpha(d,N)],\; \; \alpha(d, N)\geq 0, 
\end{equation}
holds for a dimension $d$ state $\rho$ with $N$ repetitions (queries). This deceptively simple equation is at the heart of every strategy considered in this review. Conceptually it states that any estimator is only as good as its worst-case performance which is dictated by its probability of failure, defined as a system passing a test protocol that it should fail. If this false positive probability has a functional dependence $\alpha(d,N)$ that grows in $N$ and does not vanish asymptotically in $d$, for example typically $\alpha(N,d)=O(1)N$ is dimension free, then failure is exponentially unlikely for all targets of the protocol and it is deemed efficient. This concept is schematically depicted in Figure \ref{binary_fig1}, where the probability that the target state $\rho$ contains the property $A$ builds exponentially fast with the number of questions $Q_k$ that are asked to repeated copies of $\rho$.

Conventionally, verification problems are distinguished from estimation problems. In past years there is a however an opposing trend attempting to integrate both into a unified, information-theoretic framework \cite{Eisert2020, hangleiter2020sampling}. 
In this respect, every partial tomography task (on finite-dimensional systems) may be posed in the decision theory point of view introduced here. To clarify this point, consider verification of certain property (e.g. presence of entanglement), the sampling complexity depends only on the required confidence $1-\delta$, typically $O(\log\delta^{-1})$ samples is required. On the other hand, we shall consider shadow-tomography like tasks where typically one is interested in estimation of mean values of certain set $A_1,\dots, A_M$ observables \cite{aaronson2017}. To embed this problem into the decision procedure one fixes the confidence $1-\delta$ and error $\epsilon$ and poses the estimation as a yes or no procedure: given a set of observables $A_1,\dots, A_M$, do their mean values lie within an $\epsilon$ interval from some (estimated) value? The set of queries $Q_k$ is adapted to encompass the set of inequalities $|\expval{A_{n}} - \expval{A_{n,e}}| < \epsilon$, with $\expval{A_{n}}$ being the ground truth and $\expval{A_{n,e}} $ the estimated value. Assuming a good estimator, if we have preset the error value $\epsilon$ and confidence $1-\delta$, then the procedure returns a binary outcome together with the set of estimates $\{...,\expval{A_{n,e}},...\}$.  The sampling complexity ranges from $O(\log M\epsilon^{-2}\log\delta^{-1})$ for protocols such as those engendered by shadow tomography to $O(d^2\epsilon^{-2}\log\delta^{-1})$ samples required for full state tomography (see Figure \ref{binary_fig1} to the right). Thus verification and estimation in this framework can be put on equal footing with the main difference being the inputs to the protocol (confidence $1-\delta$ for verification VS confidence $1-\delta$ and error $\epsilon$ for estimation) and their respective outputs (estimation procedure returns the set of estimates in addition to the binary yes/no output).

In a similar spirit, we require this demarcation not just in time but space as well, insisting on simple-to-implement queries on each quantum state. This will almost always mean local queries on the target system alone, rather than for example global (entangled) measurements on multiple instances. Finally, the computation of the decision function $S(\mathbf{q},\mathbf{i})$ itself must also be efficient, in that it cannot have a computational complexity that depends on the system dimension in any significant way. To summarise our requirements:
\begin{enumerate}
    \item{Dimension demarcation: $\alpha(d,N)$ is not asymptotically small for large $d$, for example $\alpha(d,N)=O(1)N$.}
    \item{Fast convergence in the number of queries: $\alpha(d,N)$ grows with $N$ for example, typically linearly.}
    \item{Low computational complexity where the measurement queries $Q_k$ are implemented by local measurement or low-depth quantum circuits.}
    \item{Simple post-processing, e.g. simple evaluation of the decision function $S(\mathbf{q},\mathbf{i})$.}
\end{enumerate}

This review will progress through query/answer strategies that satisfy these demanding properties in the following way. Section 2.1 constructs an explicit probabilistic detection scheme in keeping with the above framework. Section 2.2 considers what tasks may be performed using this protocol with the minimum access to a quantum state, converging on an entanglement verification protocol that uses only a single copy of a quantum state. Section 2.3 relaxes the single copy regime to that of dozens, observing the increase in information extraction possible in an experimental setting. Section 2.4 gives a brief summation of related works, accentuating the extension of our method to quantum state verification and certification. Section 3.1 considers the limit of the few-copy regime, considering the maximal amount of information one can extract from any quantum state, of any size, given a fixed number of samples.
Finally, Section 4 contains a recapitulation of all important points, addressing works that go beyond techniques mentioned in the review and discusses open questions.

\section{Entanglement verification}
\label{singlecopy}

In searching for worthwhile tasks, it is not a contentious statement that entanglement represents a crucial resource in many quantum-information protocols \cite{horodecki2009quantum}.
For this reason, the task of entanglement verification has by necessity spurred the development of a variety of different approaches over the past years \cite{friis2019entanglement}. Traditionally, the methods of detection (see \cite{friis2019entanglement, guhne2009entanglement} for a focused review) rely on the estimation of expectation values of observables linked to certain fundamental inequalities, such as is the case of entanglement witnesses~\cite{guhne2009entanglement, terhal2000bell,bruss2002reflections}, Bell inequalities~\cite{baccari2017efficient,rabelo2011device,werner2001bell} or the use of quantum Fisher information~\cite{akbari2019entanglement,hyllus2012fisher,li2013entanglement}, local uncertainty relations \cite{zhao2019entanglement} and nonlinear witnesses \cite{guhne2006nonlinear}.

Typically, strategies will involve testing if (some function of) the expectation value(s) of some observable(s) exceeds a certain threshold, such as testing if $\expval{W}<W_c$ and demanding, in practice, repeated measurements on large ensembles of identically prepared copies. This can be costly in terms of experimental requirements, scaling to impossibility with just a few steps as in photonic systems where coincidence rates fall exponentially fast in the system size \cite{adcock2018hard}.  
An impressive yet example of this may be found in a recent 12-photon entanglement witness experiment \cite{zhong201812}, where the detection rate was approximately one copy per hour. The extraction of a mean value of a single local observable, which typically requires one hundred to one thousand copies of a given quantum state, in this case, translated to an experiment duration measured in weeks. 
Such non-viability is a consequence of the indirect approach for testing entanglement. If instead we employ the direct method in which we pose the detection question differently, i.e., to ask: "What is the chance for the system to achieve a value $W<W_c$ in a single-shot experiment?'', we can gain a vast reduction in the detection complexity.
In this respect, we will review several highly efficient methods \cite{dimic2018single, saggio2019experimental, zhu2019optimal} based on the information-theoretic framework introduced in the previous section.

\subsection{Probabilistic detection scheme}

Consider a quantum system consisting of $n$ subsystems, each residing in a finite-dimensional Hilbert space of dimension $d$. 
The first step in any partial tomography is to define the relevant set of queries $Q_m$ that will be used to interrogate the system -- as no information may be gleaned without them. Commonly these correspond to certain binary local measurements associated to yes/no questions. For the sake of generality, we include here the quantum measurements that go beyond binary logic, that is, the positive-operator valued measures (POVMs) $E^{(k)}_{i|m}$, where $\sum_{i}E^{(k)}_{i|m}=\mathbb{1}^{(k)}$. Here $k$ labels the subsystem, $m\in\{1,...,L\}$ the local measurement setting, and $i$ is the measurement outcome. For every subsystem, we can generate one random query associated to the setting $m_k$ which when applied to the $k$th party results in some outcome $i_k$.
\begin{figure*}[ht!]
\centering
\includegraphics[width=13cm]{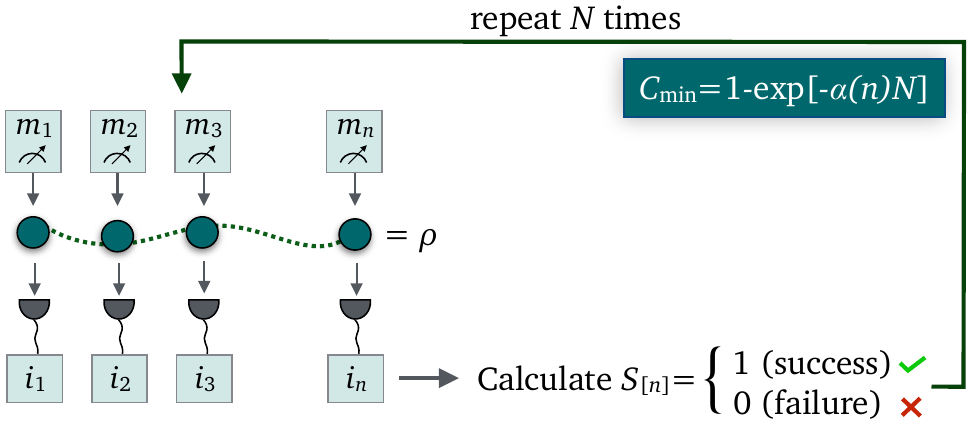}
\caption{\textbf{Probabilistic entanglement detection}. 
A single copy of an $n$-partite quantum system $\rho$ is repeatedly interrogated via random (local) measurements $m_1$,$m_2$,...,$m_n$. The performance of the system is measured via the evaluation of a cost function $S_{[n]}$. Repeating this procedure $N$ times, the probability of detecting entanglement goes to unity exponentially fast in $N$ for target state preparations, i.e., the (lower bound on) detection confidence grows as $C_{\min}=1-\exp[-\alpha(n)N]$.}
\label{singlefew}
\end{figure*}

The probabilistic entanglement detection procedure, schematically shown in Figure \ref{singlefew}, goes as follows:
\begin{itemize}
  \item[1.] A sequence of random local measurements $(m_1,...,m_n)$ drawn from a prior distribution $\Pi(m_1,...,m_n)$ is applied to a copy of quantum state $\rho$ to generate the sequence of outcomes $(i_1,...,i_n)$. 
  \item[2.] A certain binary cost function of settings and outcomes $S_{[n]}=S_{m_1...m_n}^{i_1...i_n}\in\{0,1\}$ 
  is calculated. 
  \item[3.] If $S_{[n]}=0/1$ we associate ``success/failure'' to the experimental run.
  \item[4.] Repeat $N$ times steps $1-3$.
\end{itemize}

The figure of merit for entanglement detection is the probability of success $P[S_{[n]}=1]$.
In essence, the cost functions are created such that this probability vanishes exponentially fast in the size of the system $n$ and/or in the number of repetitions $N$ for all separable states $\rho_{sep}$:
\begin{equation}\label{sepbound}
P_{\rho_{sep}}[S_{[n]}=1]\leq\exp[-\alpha(n)N],
\end{equation}
where $\alpha(n)$ is a function depending on the particular strategy and system's size. On the other hand, the procedure is tailored to detect entanglement in the vicinity of some target state $\rho_T$, i.e., $P_{\rho_T}[S_{[n]}=1]\approx 1$, thus, given the target-state preparations and desired detection confidence $1-\delta$, we can estimate the average number of copies required to verify entanglement:
\begin{equation}\label{few_copy_log_scaling}
N=\frac{\log\delta^{-1}}{\alpha(n)}.
\end{equation}

It is abundantly clear that as long as $\alpha(n)$ is not vanishingly small with the size $n$, for example, $\alpha(n)=O(1)$, we will have a logarithmic growth of the number of copies in $\delta$. Considering it in the opposite direction: the confidence for entanglement detection grows exponentially fast in the number of repetitions $N$ which constitutes what we dub the few-copy detection regime \cite{saggio2019experimental} where we achieve the high confidence detection by measuring only (thus the name) a few copies of the system (see Section 2.3). 

The reduction of resources can be further traced down in the case where $\alpha(n)$ grows in $n$. In this case, for a sufficiently large system (large $n$) this number is reduced to the logical minimum leading to the single-copy detection \cite{dimic2018single,dimic2018central}. This possibility is presented in detail in the next section.

An important aspect of these methods is that they bypass the so-called i.i.d. (independent and identically distributed) assumption taken for granted in standard approaches. This assumption means that a source produces identical copies of a quantum state in every experimental run. This is very questionable from a practical point of view, especially given the lack of perfect control and manipulation as is the case for NISQ systems. In contrast, the shown methods surpass i.i.d. through use of random sampling a set of measurement queries. In this case, the entanglement is seen as the ability of a system to compute a certain cost function (as quantified by the probability of success) in a single-shot experiment. In such a construction of the problem, the i.i.d. requirement may be relaxed without compromising the protocol. 

\subsection{Single-copy scenario}

We review the construction of the single-copy detection procedure for $k$-producible states~\cite{guhne2005multipartite} which naturally extends to cluster states~\cite{briegel2001persistent}. 
Further examples include ground states for local Hamiltonians with the entanglement gap~\cite{dowling2004energy}, among which we find many important classes of quantum states, such as the matrix product states~\cite{perez2006matrix} and projected-entangled pair states~\cite{verstraete2008matrix}.
In all examples provided we explicitly constrain to a single experimental repetition ($N=1$) and attempt to optimise the chance of entanglement detection. We put the main emphasis on the construction of protocol, i.e., appropriate choice of the settings and cost function.

\subsubsection{Example of \texorpdfstring{$\boldsymbol{k}$}{Lg}-producible quantum state}
We start with the example of the $k$-producible entangled state~\cite{guhne2005multipartite}, i.e., $\ket{\phi_1}\ket{\phi_2}\dots\ket{\phi_m}$,
where the products $\ket{\phi_s}$ involve at most $k$ parties.\footnote{This example is rather explanatory and used to demonstrate the method. A ``real'' example of cluster states will naturally follow in the next section.} Our aim is to show that entanglement can be detected with one copy of an $n$-folded state as long as $n$ is large. To clarify the probabilistic procedure even better, we take the target state to be the product of quantum singlets $\ket{\psi_0}=\ket{\psi^{-}}^{\otimes n}$,
where $\ket{\psi^{-}}=\frac{1}{\sqrt2}(\ket{01}-\ket{10})$.
The quantum singlet has the property of being the only state that returns perfect anti-correlations (the outcome $-1$) when measured with one of the operators $X\otimes X$, $Y\otimes Y$, or $Z\otimes Z$. Therefore, the suitable measurements to identify singlet uniquely are the following projectors

\begin{equation} \label{singletprojectors XYZ}
Q_X=\frac{\mathbb{1}-X\otimes X}{2},\;\; Q_Y=\frac{\mathbb{1}-Y\otimes Y}{2}, \;\; Q_Z=\frac{\mathbb{1}-Z\otimes Z}{2}.
\end{equation}

The pertinent fact is that no separable state may reveal $Q_X=Q_Y=Q_Z=1$ simultaneously; as already emphasised, this is the unique property of the target singlet state. Thus, the maximum probability to obtain the outcome $1$ for all separable inputs if measurement settings  are uniformly sampled from the set $\{XX,YY,ZZ\}$ is $2/3$:
\begin{equation}\label{singlet Fmean}
{P}_{\rho_{sep}}=\avg{\frac{1}{3}(Q_X+Q_Y+Q_Z)}\leq\frac{2}{3},
\end{equation}
for all separable two-qubit states $\rho_{sep}$. 
With this we can construct detection procedure for $n$ pairs as follows: the set of $2n$ qubits is divided into consecutive pairs and for each pair, a random measurement from the set $\{XX,YY,ZZ\}$ is applied to get a sequence of results $...,(i_k,j_k),...$. From these measurement outcomes we construct the following local cost function for every pair $S_k=\frac{1}{2}\left(1-(-1)^{i_{k}+j_{k}}\right)$, where $k=1...n$ labels the qubit pair.
Now, given bound \eqref{singlet Fmean}, the relative frequency of the outcome $1$ shall not exceed $2/3$ significantly for all separable states. Therefore, we define the overall test to be
\begin{equation}
S_{[n]}=\left\{
          \begin{array}{ll}
            1, & \hbox{$\sum_{k=1}^{n}S_k\geq (\frac{2}{3}+\epsilon)n$;} \\
            0, & \hbox{$\sum_{k=1}^{n}S_k< (\frac{2}{3}+\epsilon)n$,}
          \end{array}
        \right.
\end{equation}
where $\epsilon>0$ is a free parameter. The overall probability of success reads
\begin{equation}\label{singlet Ps}
P[S_{[n]}=1]=P\left[S_{1}+\dots+S_n\geq \left(\frac{2}{3}+\epsilon\right) n\right].
\end{equation}

Using the standard Chernoff bound~\cite{chernoff1952measure} we obtain:
\begin{equation}\label{singletsepbound}
P_{\rho_{prod}}[S_{[n]}=1]\leq e^{-D(\frac{2}{3}+\epsilon||\frac{2}{3})n},
\end{equation}
where $D(x||y)=x\log{\frac{x}{y}}+(1-x)\log{\frac{1-x}{1-y}}\geq0$
is the Kullback--Leibler divergence.
The probability of success vanishes exponentially fast in $n$ for all $\epsilon>0$.
The procedure is convenient as we do not have to set $\epsilon$ in advance, i.e., we calculate $\epsilon$  as the experimental deviation of the measured sum $\frac{1}{n}\sum_{k=1}^n S_k$ from the separable bound $2/3$.

In the perfect case of $n$ singlets $\ket{\psi_0}=\ket{\psi^{-}}^{\otimes n}$, we shall measure $S_k=1$ deterministically, thus we find that $\epsilon=1/3$.
 The bound \eqref{singletsepbound} becomes
\begin{equation} \label{singletsinglet}
P_{\rho_{sep}}[S_{[n]}=1]\leq \left(\frac{2}{3}\right)^n.
\end{equation}
Therefore, if $n$ is large enough, a single copy of $\ket{\psi_0}$ is sufficient to certify entanglement with high probability. For example, already for $n=8$, the confidence level for entanglement detection is at least $96\%$. 

Before we proceed further, let us illustrate the i.i.d. issue in following situation. Suppose that we have only $n=8$ qubit pairs at our disposal and we want to inspect the presence of entanglement. Given the prescription above, we may try to measure the witness operator $W=\frac{1}{3}(Q_X+Q_Y+Q_Z)$. However, it is not clear how to divide 8 pairs into three groups to measure three local observables $Q_X$, $Q_Y$ and $Q_Z$. Also, there is no guarantee for these pairs to be in an i.i.d. state $\rho_{12}^{\otimes 8}$ which seems to be needed for separate estimation of $\avg{Q_X}$, $\avg{Q_Y}$ and $\avg{Q_Z}$. In this case, it is not clear how to proceed. For example, we may use the first three copies to measure $Q_X$, the second three to measure $Q_Y$, and the last two for the measurement of $Q_Z$. However, if the order of measurements is known in advance we may arrive at false entanglement verification: the following product state $\ket{\phi_p}=(\ket{x+}\ket{x-})^{\otimes3}(\ket{y+}\ket{y-})^{\otimes3}(\ket{z+}\ket{z-})^{\otimes2}$ gives exactly the same result as the i.i.d. singlet state $\ket{\psi^{-}}^{\otimes 8}$ for these fixed measurements. The key procedure to surpass i.i.d. assumption is random sampling and the probabilistic detection described above. It provides a clear separation between the state $\ket{\psi^{-}}^{\otimes 8}$ and the product state $\ket{\phi_p}$, as the later has only the chance of $(2/3)^8\approx0.039$ in the best case to reveal the result $S_{1}+\dots+S_8=8$. In contrast, the experiment with the single-state preparation $\ket{\psi_0}$ reveals ``success'' always thus we verify entanglement with at least $C_{\min}=1-0.039\approx0.96$ confidence.

\subsubsection{Single-copy detection of cluster states} \label{ECS}

Another example we present here is that of cluster states~\cite{briegel2001persistent} as a natural generalisation of the previous example of $k$-producible state. In contrast however, cluster states contain genuine multiparty entanglement~\cite{hein2004multiparty} and represent a universal resource for measurement-based quantum computation~\cite{raussendorf2001one}. For simplicity, we work out in detail an example of a linear cluster state (LCS); generalisations of the scheme to higher dimensions are straightforward and briefly discussed at the end of the section. 
 
The $n$-qubit LCS is uniquely defined by the set of $2^n$ stabilizers
\begin{equation}\label{stab equation}
G_{q_1\dots q_n}\ket{LCS}=G_1^{q_1}\dots G_n^{q_n}\ket{LCS}=+1\ket{LCS},
\end{equation}
where $G_k=Z_{k-1}X_kZ_{k+1}$ and $q_k=0,1$. Here $\{X_k,Y_k,Z_k\}$ is the set of standard Pauli matrices acting on $k$th qubit and without loss of generality we have chosen the cluster state with periodic boundaries, i.e., $Z_{n+1}\stackrel{\mathrm{def}}{=}Z_1$ and $X_{n+1}\stackrel{\mathrm{def}}{=}X_1$. 

Let us analyse a small sub-cluster of four qubits (e.g. qubits $\{1,2,3,4\}$) with the corresponding stabilizers 
\begin{equation} G_2=Z_1X_2Z_3, \; \; G_3=Z_2X_3Z_4 \;\; \text{and} \;\; G_2G_3=Z_1Y_2Y_3Z_4
\end{equation}
acting exclusively on it. Even though these three stabilizers are commutative, they are not locally compatible, which means one can not measure all three simultaneously with local measurement. Therefore, there is no separable state for which $G_2=G_3=G_2G_3=+1$ simultaneously.
Consequently, if we randomly chose to measure one of the stabilizers (with probability $1/3$), there is only a chance of $2/3$ to get the result $+1$, for all separable inputs. This observation empowers our detection method to work. 
The strategy is to pick a random partition of the set of $n$ qubits into $4$-qubit clusters and then measure one of the corresponding stabilizers randomly on each of them. Given our previous analysis, the relative frequency of the outcome $+ 1$ can not substantially surpass the value of $2/3$. It is convenient to introduce regular partitions (i.e., neighbouring clusters overlap on at most one qubit) of $n$-qubit cluster state into $L$-partition of 4-qubit clusters $\{c_{t_1},c_{t_2},\dots c_{t_L}\}$, where $c_{t_k}$ is the cluster consisting of the sequence of four neighbouring qubits:
\begin{equation}
c_{t_k}=\{t_k,t_k+1,t_k+2,t_k+3\}. 
\end{equation}
The set of all regular partitions of size $L$ is denoted by $\mathcal{C}_L$.

For every cluster $c_{t_k}$ in the partition we associate three stabilizers:
 \begin{equation}
 \begin{split}
 & G_{t_k+1}=Z_{t_k}X_{t_k+1}Z_{t_k+2},\; \\
  & G_{t_k+2}=Z_{t_k+1}X_{t_k+2}Z_{t_k+3} \;  \text{, and} \\
\; & G_{t_k+1,t_k+2}=G_{t_k+1}G_{t_k+2}=Z_{t_k}Y_{t_k+1}Y_{t_k+2}Z_{t_k+3}.
\end{split}
\end{equation}
To each of them we associate three projectors
\begin{equation}
Q_{t_k}=\frac{\mathbb{1}+G_{t_k+1}}{2},~W_{t_k}=\frac{\mathbb{1}+G_{t_k+2}}{2},~R_{t_k}=\frac{\mathbb{1}+G_{t_k+1}G_{t_k+2}}{2},
\end{equation}
projecting on the $+1$ outcome. To these we associate the following measurement settings $\{ZXZZ,ZZXZ,ZYYZ\}$, and we assign ``success'' to the cluster measurement only if the outcome $+1$ (for the value of measured stabilizer) occurs. Formally speaking, for every cluster we define the following local cost function
\begin{equation}\label{local cost}
S_k=S_{m}^{i_1i_2i_3i_4}=\frac{1}{2}+\frac{1}{2}\left\{
                           \begin{array}{ll}
                             (-1)^{i_1+i_2+i_3}, & \hbox{$m=ZXZZ$;} \\
                             (-1)^{i_2+i_3+i_4}, & \hbox{$m=ZZXZ$;} \\
                             (-1)^{i_1+i_2+i_3+i_4}, & \hbox{$m=ZYYZ$,}
                           \end{array}
                         \right.
\end{equation}
where $k=1\dots L$. Finally, for a given partition $\{c_{t_1},c_{t_2},\dots,c_{t_L}\}$ the overall cost function is represented in the following way
\begin{equation}\label{overall cost}
S_{[n]}=\left\{
          \begin{array}{ll}
            1, & \hbox{$S_1+\dots+S_L\geq (\frac{2}{3}+\epsilon)L$;} \\
            0, & \hbox{$S_1+\dots+S_L< (\frac{2}{3}+\epsilon)L$,}
          \end{array}
        \right.
\end{equation}
where $\epsilon>0$ is a free parameter. We associate ``success'' to the experimental run if the number of local successes exceeds a certain threshold of $(\frac{2}{3}+\epsilon)L$.
The detection procedure goes as follows:

{\begin{itemize}
\item[1.] Randomly generate a partition of $n$-qubit cluster state $\{c_{t_1},c_{t_2},\dots,c_{t_L}\}$ from the set $\mathcal{C}_L$
with probability $1/|\mathcal{C}_L|$.
\item[2.] Draw one measurement setting for each cluster in the partition with probability $1/3$.
\item[3.] Perform local measurements and collect the sequence of results  $S_1,S_2,\dots,S_L$.
\item[4.] Calculate the cost function $S_{[n]}$ by using \eqref{overall cost}.
\end{itemize}}

We shall analyse the probability to pass the test for separable states. Firstly, for all product states the local cost functions $S_k$ are independent binary random variables with $\avg{S_k}\leq2/3$ for all $k=1\dots L$. The overall probability of success reads
\begin{equation}
{P}_{\rho_{prod}}[S_{[n]}=1]=P_{\rho_{prod}}\left[S_{1}+\dots+S_L\geq \left(\frac{2}{3}+\epsilon\right) L\right],
\end{equation}
which is the probability that the sum of independent random variables $S_1+\dots+S_L$ exceeds the value of $(\frac{2}{3}+\epsilon) L$. 
As $\avg{S_k}\leq2/3$, the sum $S_1+\dots+S_L$ cannot exceed $2/3L$ significantly. Indeed, as before, the Chernoff bound holds (for detailed proof see Supplementary Information of \cite{dimic2018single}), i.e.,
\begin{equation}\label{LCSsepbound}
{P}_{\rho_{prod}}[S_{[n]}=1]\leq e^{-D(\frac{2}{3}+\epsilon||\frac{2}{3})L}
\end{equation}
where $D(x||y)$ is the Kullback-Leibler divergence. As the bound holds for all product states, it also holds for their mixtures, i.e., for all separable states.

On the other hand, for the case of cluster state preparation $\ket{LCS}$, each local cost function takes the value $S_k=1$, thus we have $\epsilon=1/3$. The bound \eqref{LCSsepbound} reduces to
\begin{equation}\label{LCSsepbound1}
P_{\rho_{sep}}[S_{[n]}=1]\leq \left(\frac{2}{3}\right)^L.
\end{equation}
For the sufficiently large number of qubits even a single-copy of the LCS suffices to certify the presence of entanglement with high probability. For example, already for $n=24$, we have $L=8$ which gives a confidence level greater than $95\%$. {Finally, let us comment briefly on the generalization to the higher dimensional case. In the case of a 2D cluster state, one can introduce partitions into $4\times4$ qubit clusters with the corresponding stabilizer projectors (using complete analogy to $Q_{t_k}$, $W_{t_k}$ and $R_{t_k}$ for LCS) and define the local cost functions. The 2D detection scheme also consists of drawing a random partition followed by a random measurement of local projectors on individual clusters. The separable bound similar to \eqref{LCSsepbound} can be derived. On the other hand, if the 2D cluster state is the input state, the probability of success is $1$.}

\subsubsection{Single-copy detection of ground-states of local Hamiltonians}

One of the strong reasons why the single-copy entanglement scheme works for the cluster states is the robustness of entanglement to local perturbations, meaning that local measurements on qubits do not destroy the entanglement between the remaining qubits completely. Thus one can expect other classes of states sharing this property to admit single-copy entanglement detection. The ground states of local Hamiltonians share this property~\cite{eldar2017local}; therefore they are good candidates.
Let us consider a $L$-local Hamiltonian on some graph of $n$ particles  $H=\sum_{k=1}^{n}H^{(k)}$, where $H^{(k)}$ acts on at most $L$ subsystems ($L$ is fixed and independent of $n$). 
Now, let $\ket{\psi_0}$ be the ground state of the Hamiltonian $H\ket{\psi_0}=n\epsilon_0\ket{\psi_0}$, where $E_0=n\epsilon_0$ is the ground-state energy.  We are working with Hamiltonians that exhibit the so-called entanglement gap
$g_{E}=\epsilon_{sep}-\epsilon_0>0$,
where $\epsilon_{sep}=\frac{1}{n}\min_{\rho_{sep}}\mathrm{Tr} H\rho_{sep}$
is the minimal obtainable energy per particle by a separable state ~\cite{dowling2004energy}. 
The main idea of the procedure is to use mean energy $\avg{H}$ as an entanglement witness: $\avg{H}\geq n\epsilon_{sep}$ holds for all separable states, while at least the ground state violates this bound. This fact can be exploited to develop an efficient probabilistic procedure by employing a tomographically complete set of measurements. In this case, the operator $H$ translates into a classical random variable $H_{[n]}$ which serves to witness entanglement in practice (the general procedure is explained in detail in the next Section 2.3). The central object for our detection protocol is then the following overall cost function:

\begin{equation}\label{Ham cost}
S_{[n]}=\left\{
          \begin{array}{ll}
            1, & \hbox{$H_{[n]}\leq n(\epsilon_{sep}-\delta)$;} \\
            0, & \hbox{$H_{[n]}> n(\epsilon_s-\delta)$,}
          \end{array}
        \right.
\end{equation}

where $0<\delta<\epsilon_{sep}-\epsilon_0=g_{E}$ is a free parameter. 
Since $\avg{H}\geq n\epsilon_{sep}$ holds for all separable states, for the case of $n$ being large, $H_{ [n]}$ is unlikely to precede the separable bound $n\epsilon_{sep}$ in a single-shot experiment. Indeed, analogously to the previous two examples, one can derive the Chernoff bound for all separable states:
\begin{eqnarray}\label{Hamsepbound}
{P}_{\rho_{sep}}[S_{[n]}=1]&\leq&\exp\left[-n\kappa^2\delta^2\right],
\end{eqnarray}
where $\kappa>0$ is constant. 
Thus, for all separable inputs, the probability of success vanishes exponentially fast with $n$. In contrast, for the ground-state preparation $\ket{\psi_0}$, the probability of success reaches $1$ in the thermodynamic limit, as it follows from the following bound:
\begin{equation}\label{GS bound}
{P}_{\psi_0}[S_{[n]}=1]\geq1-\frac{\beta^2}{n(g_{E}-\delta)^2},
\end{equation}
where $\beta>0$ is constant.
The first inequality \eqref{Hamsepbound} is the consequence of the McDiarmid's inequality, while the second \eqref{GS bound} is derived by using the Chebyshev's inequality. Both bounds are rigorously derived in the Supplementary Information of Ref.  \cite{dimic2018single}.

\subsubsection{Tolerance to noise}
In the end, we briefly comment on the effects of noise on single-copy entanglement detection. Consider a $n$-partite target state $\rho_0$ which passes the single-copy test with probability $p_0$.
In practice, one needs on average $1/p_0$ copies of $\rho_0$ to detect entanglement. On the other hand, let the separable bound hold, meaning that the probability of success for all separable inputs is exponentially small in $n$. We consider a mixture $\rho=\lambda\rho_{sep}+(1-\lambda)\rho_0$, 
where $\rho_{sep}$ is an arbitrary separable state and parameter $0<\lambda<1$ quantifies the amount of noise.
The overall probability of success is a mixture of probabilities
$P_{\rho}=\lambda P_{\rho_{sep}}+(1-\lambda)P_{\rho_0}\approx(1-\lambda)p_0$, as long as $(1-\lambda)p_0$ is significantly larger than $P_{\rho_{sep}}=O(\exp[-nc])$. This implies that noise impacts detection by suppressing the probability of success by a factor $1-\lambda$, for any kind of noise representable by a separable state. Therefore, one requires on average $\frac{1}{(1-\lambda)p_0}$ experimental runs to confirm the presence of entanglement. This represents a strong resistance to noise as long as $(1-\lambda)p_0$ is not exponentially small in $n$. For example, if we consider $(1-\lambda)p_0>0$ constant and independent of $n$, then we verify entanglement with a fixed cost in terms of the number of samples. This described scenario is very different in comparison with conventional detection techniques. Generally, a witnessing method tolerates noise below a certain critical point, i.e., $\lambda<\lambda_c$, meaning that if noise passes the threshold, the scheme fails to detect entanglement.

\subsection{Entanglement detection with a few copies}\label{fewcopies}

In this section we review an entanglement detection method where the required number of copies grows logarithmically slow with the confidence as shown in equation \eqref{few_copy_log_scaling}. The main goal of this section is to translate one of the most common methods for entanglement detection, that is, the one based on entanglement witnesses~\cite{terhal2000bell,bruss2002reflections} (see \cite{guhne2009entanglement} for concise review), into an efficient framework that requires only a few experimental repetitions. 

What makes the witness-based technique practical is the simplicity of its detection criterion, based on a simple mean value estimation of a single (witness) observable.
Specifically, an observable $W$ is designated a witness if $\avg{W}=\mathrm{Tr}(W \rho_{\mathrm{sep}})\geq0$ for all separable states $\rho_{\mathrm{sep}}$, while $\avg{W}<0$ holds for at least one entangled state.  
In principle, we can construct an entanglement witness for every entangled state $\rho$ (theorem of completeness of witnesses~\cite{horodecki2001separability}), which is then used to detect entanglement in a target state.  
While straightforward, a drawback of the method is that the witness $W$ cannot be accessed locally, instead it must be decomposed into a sum of local observables $W=\sum_{i=1}^L W_i$ that must be individually estimated. This means that the mean value $\avg{W}$ is obtained from the $\avg{W_i}$'s, each of which is measured in an independent experiment. The sampling complexity of the procedure is therefore dependent on the number of local terms $L$, which become a significant factor for generic witnesses on a large system. To overcome this problem, remarkable effort has been put into constructing entanglement witnesses whose measurement requires a smaller number of measurement settings, thus reducing the experimental requirements \cite{laskowski2013optimized,knips2016multipartite,chen2020verification,bavaresco2018measurements} (for more references, see recent review \cite{friis2019entanglement}). For example, refs.~\cite{guhne2002detection,lu2007experimental} find optimal decompositions of entanglement witnesses into a few local operators, even reducing  in some cases the witness decomposition to only two local operators~\cite{toth2005detecting}. However, even with a minimal number of measurement settings, this method may become inconvenient or even unfeasible simply due to the lack of sufficient number of copies of the resource state needed to extract the witness expectation value.
In such cases, alternative methods going beyond mean-value extraction are required. We review here the general method developed in Ref.~\cite{saggio2019experimental} that translates the witness method into a resource-efficient probabilistic framework described in Section 2.1. In this scenario, the typical procedure achieves very high confidence in entanglement detection with just few experimental repetitions (copies of target state). As we shall see, the number of measurement settings involved into the local decomposition is not the crucial parameter determining the sampling complexity, in contrast to the standard belief~\cite{toth2005detecting}. We also review an experiment performed with a photonic system to test the practicality of the method ~\cite{saggio2022perspective}.

\subsubsection{Embedding entanglement witnesses in a probabilistic detection framework}
The aim of this section is to review the translation of any entanglement witness into the probabilistic framework. 
\label{tool}
As previously discussed, an entanglement witness $W$ is normalised such that
\begin{eqnarray}
&\ \ \avg{W}_{\mathrm{s}}=\mathrm{Tr}(W \rho_{\mathrm{s}} ) \geq 0  
    \label{witness_condition}
\end{eqnarray}
for all separable states $\rho_{\mathrm{s}}$. 
On the other hand there exists at least one entangled state $\rho$ for which $\avg{W}=\mathrm{Tr}(W \rho) < 0$.
The witness operator is normally tailored to detect entanglement in the vicinity of some target state for which $\avg{W}$ reaches the lowest possible value. We shall slightly change the general form of $W$ and introduce the witness operator $O$ in the following way: 
\begin{equation}
W=\gamma_{\mathrm{s}} \mathbb{1}-O,
    \label{witness}
\end{equation}
%the separability condition
thus equation\ \!\eqref{witness_condition} translates to
\begin{equation}
    \avg{O}_{\mathrm{s}} = \mathrm{Tr}(O \rho_{\mathrm{s}}) \leq \gamma_{\mathrm{s}}
    \label{ogamma}
\end{equation}
for all separable states $\rho_{\mathrm{s}}$. Now $O$ can be decomposed in terms of $L$ local observables $O_i$ as $O=\sum_{i=1}^L O_i$, where each $O_i$ can be turned into a non-negative observable by adding a constant term, i.e., $O_i^{'}=O_i + \alpha_i \mathbb{1}\geq0$ with $\alpha_i \geq 0$. Thus we get a new witness operator $O^{'}=\sum_{i=1}^L O_i^{'}= \sum_{i=1}^L (O_i + \alpha_i \mathbb{1}) = O + \alpha \mathbb{1}$, with $\alpha=\sum_i\alpha_i$, which is positive semi-definite operator. Inequality \eqref{ogamma} translates to the new condition:
\begin{equation}
    \avg{O^{'}}_{\mathrm{s}} = \avg{O}_{\mathrm{s}} + \alpha L \leq \gamma_{\mathrm{s}} + \alpha L
    \label{oprimegamma}
\end{equation}
for all separable states $\rho_{\mathrm{s}}$. We now write the spectral decomposition of $O_i^{'}$ in terms of eigenprojectors (i.e., binary observables) $M_{ik}$ as $O_i^{'}=\sum_{k=1}^{J_i} \lambda_{ik} M_{ik}$, where $\lambda_{ik}\geq0$ because $O_i$'s are non-negative. Here $J_i$ counts the non-zero eigenvalues of $O_i$. Since $O_i^{'}$ are local, $M_{ik}$ can be as well chosen to be local operators. To simplify the notation, we define the constant $\tau=\sum_{i=1}^{L} \sum_{k=1}^{J_i} \lambda_{ik}$ and we set $\mu_{ik}=\lambda_{ik}/\tau\geq0$. Finally, the witness condition \eqref{oprimegamma} reads
\begin{equation}
    \sum\limits_{i=1}^{L} \sum\limits_{k=1}^{J_i} \mu_{ik} \mathrm{Tr}(M_{ik} \rho_{\mathrm{s}}) \leq \frac{\gamma_{\mathrm{s}} + \alpha L}{\tau}=p_s,
    \label{fullp}
\end{equation}
for all separable states $\rho_s$. The last formula completely determines a probabilistic procedure for detection. Namely, since $\sum_{ik}\mu_{ik}=1$ and $\mu_{ik}\geq0$, these numbers are sampling probabilities for local binary observables $M_{ik}$. The LHS of the equation is just the probability of success to get $M_{ik}=1$, while the RHS is the corresponding separable bound $p_s$.
On the other hand, for target state preparations we have violation of separable bound \eqref{oprimegamma} which directly translates to a different probability of success (the entanglement value) $p_{\mathrm{e}} = (\gamma_{\mathrm{e}} + \alpha L)\tau $, with $\gamma_{\mathrm{e}}>\gamma_{\mathrm{s}}$ or equivalently the deviation $p_{\mathrm{e}}-p_{\mathrm{s}}>0$.
 
To summarise, the procedure consists of the following steps:
\begin{itemize}
    \item[1.] Randomly measure observables $M_{ik}$ (with probability $\mu_{ik}$) $N$ times to get the sequence of results $m_1,...,m_N$;
    \item[2.] Calculate the observed success rate $S_{[N]}=\frac{1}{N}(m_1+...+m_N)$. 
\end{itemize}
As before, we do not expect $S_{[N]}$ to significantly exceed the separable bound $p_s$ for all separable states, which is encapsulated into the following Chernoff bound 
\begin{equation}
{P_{\rho}}_{sep}[S_{[n]}\geq p_s+\epsilon] \leq e^{-D(p_s+\epsilon||p_s)N}.
    \label{sep_bound_few}
\end{equation}
On the other hand, for target state preparation we expect $S_{[N]}\approx p_e$ and thus the average number of target-state copies needed to achieve some fixed confidence $C=1-\delta$ is estimated as
\begin{equation}
N\approx\frac{\log\delta^{-1}}{D(p_e||p_s)}.\label{minconf}
\end{equation}
This number grows in a logarithmic fashion with the required confidence and as we shall see from examples below, only a few copies are needed to detect entanglement with a very high confidence.

\subsubsection{Example I: Projective witness for graph states}
\label{witnessW1}
Consider the standard projective witness for a graph state $\ket{G}$~\cite{guhne2009entanglement}:
\begin{equation}
W_1=\frac{1}{2} \mathbb{1} - \ketbra{G}{G},
    \label{W1}
\end{equation}
tailored for detection of genuine multipartite entanglement. This witness comes already in the form of \eqref{witness}, and it is therefore straightforward to identify the parameter $\gamma_{\mathrm{s}}=1/2$ and the observable $O=\ketbra{G}{G}$. We also have the local decomposition $O=\sum_{i=1}^{2^n} S_i/2^n$, where $S_i$ are stabilizers of state $\ket{G}$ and are in general tensor products of the Pauli operators~\cite{browne2016one}. 
One can therefore easily identify $L=2^n$ and $O_i=S_i/2^n$. The operators $O_i$ have to be shifted for $\alpha_i=1/2^n$ to get non-negative observables $O^{'}=\sum_{i=1}^{2^n} (S_i/2^n + \mathbb{1}/2^n)$. These are already in eigenform, thus we have $J_i=1$, $\tau=2$, $\lambda_i=2/2^n$ and $M_i=(S_i + \mathbb{1})/2$. The sampling probabilities are $1/2^n$ and the separable bound is calculated from
\begin{equation}
\sum_{i=1}^{2^n} \frac{1}{2^n} \mathrm{Tr}(M_i \rho_{\mathrm{s}}) \leq \frac{3}{4}=p_{\mathrm{s}}.
\label{pforW1}
\end{equation}
On the other hand, for the target state preparation $\rho_T=\ketbra{G}{G}$ we have
\begin{equation}
    \sum_{i=1}^{2^n}\frac{1}{2^n}  \mathrm{Tr}(M_i \rho_T) = 1,
    \label{prob_success_ent}
\end{equation}
thus the entanglement value reads $p_{\mathrm{e}}=1$. 
To estimate the number of copies, we can choose, for example, a confidence  of $1-\delta=0.99$. Equation \eqref{minconf} gives us $N\approx\mathrm{log} (1-0.99)^{-1}/D(1||3/4) \approx 16$, which is a notably small number. A naive approach of measuring all $2^n$ observables $M_i$ independently will quickly become unfeasible, while with the probabilistic detection we achieve the same confidence with a constant number of copies, regardless of the system size. 

\subsubsection{Example II: witness requiring two local measurements}
The second example we consider is the witness tailored to detect entanglement in $n$-qubit cluster state $\ket{C}$ presented in Ref.~\cite{toth2005detecting} (an equivalent example is also presented for the GHZ state which in full analogy can be adapted here). An optimal witness decomposition for detecting genuine multipartite entanglement requiring only two measurement settings is found:
\begin{equation}
W_2= 3 \mathbb{1} - 2 \Big( \prod\limits_{\mathrm{even\ } i}  \frac{\mathbb{1}+ G_i}{2}    +  \prod\limits_{\mathrm{odd\ } i} \frac{\mathbb{1}+ G_i}{2}  \Big),
    \label{witness2}
\end{equation}
with $i=1,...,n$. The observables $G_i$ are called generators of the state (in this case the cluster state $\ket{C}$), and constitute a subset of the stabilizing operators $S_i$. To translate this witness, we can apply the procedure described in Subsection 2.3.1. Firstly, we easily identify $\gamma_{\mathrm{s}}=3$ and $O=2 \Big( \prod_{\mathrm{even\ } i}  \frac{\mathbb{1}+ G_i}{2}    +  \prod_{\mathrm{odd\ } i} \frac{\mathbb{1}+ G_i}{2}  \Big)$. We notice that $O$ is already decomposed into two non-negative binary observables $M_1=\prod_{\mathrm{even\ } i}  \frac{\mathbb{1}+ G_i}{2}$ and $M_2=\prod_{\mathrm{odd\ } i} \frac{\mathbb{1}+ G_i}{2} $ and the sampling probabilities are $1/2$. The separable bound is given by
\begin{equation}
\sum_{i=1}^2 \frac{1}{2}\mathrm{Tr}(M_i \rho_{\mathrm{sep}})\leq \frac{3}{4}=p_s.
    \label{prob_success_2}
\end{equation}
On the other hand, the target state preparation returns $p_e=1$ and the estimated number of copies entirely matches the analysis provided in the previous example. From here we see that although the projective witness \eqref{W1} involves exponential terms in the local decomposition, it performs equally well as the witness with two settings only.

\subsubsection{Generic witness}

In the last two examples, the sampling complexity was completely independent of the system size: the average number of required copies solely depends on the required confidence for entanglement detection. However, we cannot expect such size-free behaviour in the general case. The key parameter dictating the scaling behaviour is the deviation between entanglement value and separable bound $p_e-p_s$, which can become asymptotically small with the size of the system. To illustrate this, we consider the example of the following witness
\begin{equation}
    W=(n-1)\mathbb{1}-\sum_{i=1}^nS_i,
\end{equation}
constructed to detect entanglement in the vicinity of the state stabilized by the set $S_1,...,S_n$ \cite{toth2005entanglement}. 
The translation procedure is very straightforward in this case resulting in the following separable bound
\begin{equation}
    \sum_{i=1}^n\frac{1}{n}\mathrm{Tr}(M_i\rho_{\mathrm{sep}})\leq1-\frac{1}{n}=p_s,
\end{equation}
where $M_i=(\mathbb{1}+S_i)/2$, while for the target state preparation we have $p_e=1$. In this case, the estimated number of copies is $N\approx\frac{\log\delta^{-1}}{D(1||1-\frac{1}{n})}$. For large $n$ this can be approximated with $N\approx n\log\delta^{-1}$, which defines a linear growth in the system size. In the general case, supposing that $\epsilon_0=p_e-p_s$ is asymptotically small in $n$, then we have two regimes: if $p_e=1$ formula \eqref{minconf} gives $N\approx\frac{\log\delta^{-1}}{\epsilon_0}$, while for $p_e<1$ this scaling becomes qudratically worse $N\approx \frac{2p_e(1-p_e)\log\delta^{-1}}{\epsilon_0^2}$. For a generic witness, as long as $\epsilon_0^{-1}=\mathrm{poly}[n]$, the procedure remains efficient.

\subsubsection{Experimental scenario}
\begin{figure*}[ht!]
\centering
\includegraphics[width=0.8\textwidth]{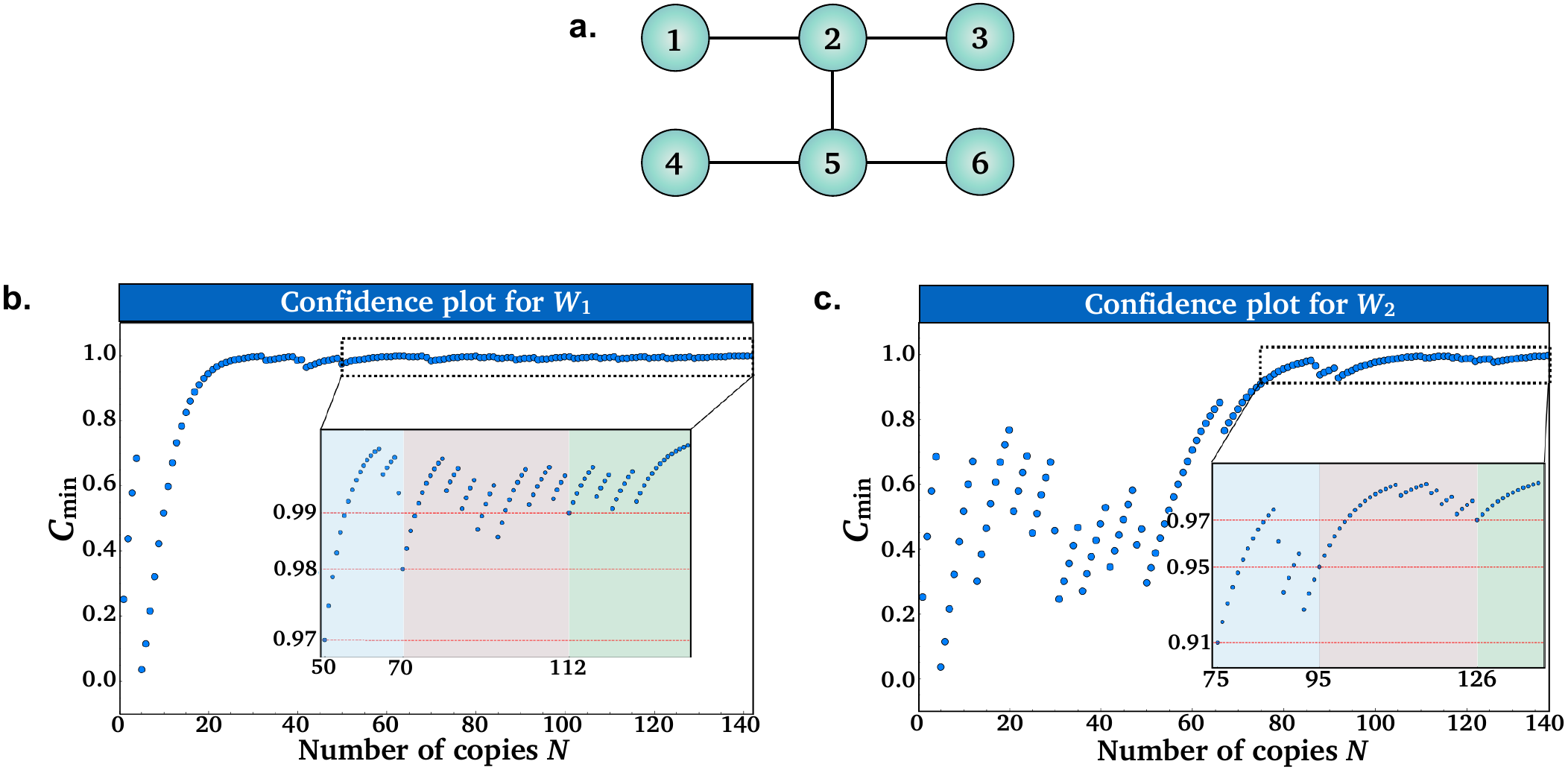}
\caption{\protect\raggedright
\textbf{Experimental scenario. (a)} H-shaped six-qubit cluster state. Each disk represents a qubit prepared in the superposition state $\ket{+}=(\ket{0}+\ket{1})\sqrt{2}$, and the solid lines indicate entanglement between them. \textbf{(b), (c)} Growth of the minimum confidence with the number of copies. The blue dots represent $C_{\mathrm{min}}$ calculated from equation~\eqref{minconf} for $W_1$ \textbf{(b)} and $W_2$ \textbf{(c)}. The insets  show the region where the confidence stabilizes. The images are adapted from Ref.~\cite{saggio2019experimental}.}
\label{experiment}
\end{figure*}

The theoretical framework presented above was tested in the experiment presented in Ref.~\cite{saggio2019experimental}. The setup was designed to produce the following six-photon cluster state 
\begin{equation}
 \begin{aligned}
%\begin{multline}
\ket{Cl_6}= &
 \frac{1}{2} (\ket{000000} + \ket{000111}+\ket{111000}-\ket{111111}),
\label{6cluster}
%\end{multline}
 \end{aligned}
\end{equation}
which is an equivalent version (up to local unitary transformations) of the six-qubit H-shaped cluster state depicted in Figure \ref{experiment}a. The state is produced with a photonic setup where logical qubits are encoded in the photons' polarization. The entanglement verification test was performed both for witnesses $W_1$ and $W_2$ introduced in equations \eqref{W1} and \eqref{witness2}. The binary observables $M_i$ defining the witness $W_1$ were sampled $N=160$ times, while the $M_i$ constituting $W_2$ were drawn $N=150$ times. The observed deviation $\epsilon=S_{[6]}-3/4$ from the separable bound was plugged into equation \eqref{sep_bound_few} to put the lower bound on the confidence for entanglement detection. Figures \ref{experiment}b, c provide the experimental plots for the two witnesses. In the case of witness $W_1$, the plot in Figure  \ref{experiment}b shows that only 50 copies of the experimental state are needed to verify genuine multipartite entanglement with at least 0.97 confidence, and that 112 suffice to reach at least 0.99. In the same way, using the witness $W_2$, it is visible from Figure \ref{experiment}c that 126 copies are enough to reach a confidence of at least 0.97. The deviation from the expected theoretical values are due to experimental imperfections that lead to a limited fidelity of $F\approx0.75$. 

\subsection{Related work}
Probabilistic detection techniques similar to those presented here can be found in several other works. In the context of Bell's inequalities, similar kinds of probabilistic protocols are constructed for the single-shot non-locality detection \cite{araujo2020bell} and entanglement detection via preparation games \cite{weilenmann2020quantum}. In the context of quantum state verification \cite{pallister2018optimal,zhang2019experimental, zhang2020experimental, liu2021universally}, a single-shot entanglement verification naturally arises in bipartite states as long as the dimension of marginal systems becomes sufficiently large \cite{zhu2019general,zhu2019efficientadversarial}. The generalisation to the GHZ states can be found in \cite{li2020optimal}. These results show a more intimate relation between our probabilistic detection and quantum state verification protocols. This is supported by the fact that our probability of success (to calculate the cost function) is usually maximised to 1 for the target state, thus the correct set of outputs does not witness only the presence of entanglement, it also indicates that the preparation state is close to the target state. Therefore, it seems that our protocols naturally extend from entanglement detection to more informative quantum state verification without significantly increasing the cost in terms of resources. Given this relation, we will review in what follows the basics of quantum state verification and its recent extension to the device-independent scenario and quantum state certification \cite{dimicsupicdakic}.

\subsubsection{Quantum state verification and certification}
The quantum state verification (QSV) is a protocol that verifies if an unknown input state is close (in fidelity) to some target state.
Due to its simplicity and low complexity, it has recently attracted a lot of attention in the community, and several verification protocols have been constructed for various classes of states \cite{zhu2019efficient,zhu2019optimal,liu2021universally,yu2019optimal,liu2019efficient,takeuchi2018verification} together with experimental demonstrations \cite{zhang2019experimental,jiang2020towards, zhang2020classical}.
From the theoretical point of view, QSV plays an important role in protocols such as blind quantum computation and quantum networks \cite{hayashi2015verifiable,fujii2017verifiable,hayashi2018self,markham2020simple,morimae2013blind, takeuchi2019quantum,perseguers2013distribution, mccutcheon2016experimental}.

In this section, we recall the framework for QSV as defined by \cite{pallister2018optimal}. The main goal is to verify if a sequence of states $S_N = \{\sigma_1,\cdots, \sigma_N\}$ is close to the target state $\sigma = \ketbra{\psi}{\psi}$ by using only local measurements. The measurement strategy labelled by $\Omega$ thus consists of $L$ different local measurements $\{M_{i|m}\}$, where $m \in \{1,\cdots, L\}$ labels the setting and $i \in \{0,1\}$ the binary outcome. In the $k$-th round a measurement from $\Omega$ is randomly sampled (with probability $p_k$) and applied to the state $\sigma_k$. We say that the state $\sigma_k$ passed the round if it returned the output $i = 1$. Otherwise, we say it failed. The first time a round is failed the process is aborted. The measurements are chosen such that the strategy operator $\hat{\Omega}=\sum_mp_mM_{1|m}$ is uniquely optimised for the target state: $\hat{\Omega}\ket{\psi}=+1\ket{\psi}$, meaning that only target state passes all verification rounds with probability 1.  
Under the premise that all emitted states are either $\bra{\psi}\sigma_k\ket{\psi}\leq1-\epsilon$ away from target state or all of them are actually target states $\sigma_k=\ketbra{\psi}{\psi}$, one can derive the average number of tests $N=\frac{\log{\delta^{-1}}}{\nu(\Omega)\epsilon}$ needed to achieve the confidence of $1-\delta$. The value $\nu(\Omega)$ is the so-called spectral gap which is the second largest eigenvalue of $\hat{\Omega}$.

The sampling complexity of the QSV is only  up to a constant factor optimal in error $\epsilon$, as the best strategy is achieved for the projection on target state measurement $\{\ketbra{\psi}{\psi}, 1- \ketbra{\psi}{\psi}\}$ resulting in $\sim\frac{\log{\delta^{-1}}}{\epsilon}$ scaling. While this is a remarkable result, the downside of the QSV scheme as proposed by \cite{pallister2018optimal} is its impracticality, i.e., the verification condition of all states either being $1-\epsilon$ away from the target or all being target states.  Such assumption is very hard to justify operationally and extremely hard to achieve in laboratory \cite{jiang2020towards}. 
In our recent work, we relax this assumption and we fully adapt the protocol to device-independent (DI) quantum state verification \cite{dimicsupicdakic}. In this case, all devices are not characterised nor trusted and all operations are treated as black-boxes \cite{colbeck,pironio2010random, diqkd,scarani}. Remarkably, we have shown that the optimal scaling of $N=O(\frac{\log{\delta^{-1}}}{\epsilon})$ translates to the DI scenario. The scheme is more practical as it tolerates $O(\epsilon)$ failure events during the verification process without losing the optimal scaling.

A general drawback of QSV is that the verification process destroys the quantum resource and the conclusion is made about the resource which is fully consumed. This prevents the possibility of using it for other protocols and further processing. The solution to this problem is found in quantum state certification: a protocol in which a fragment of the resource copies is measured to authorise the rest of the copies. The pioneering quantum state certification protocols are developed in \cite{zhu2019general, zhu2019efficientadversarial}. In these works, one explores permutation symmetry and measures all but one copy, which then serves as a certificate. The protocol is very powerful as it applies to a generic adversarial scenario, but it unfortunately consumes $O(N)$ resources to certify a single copy only. Our new approach on DI QSV developed in Ref. \cite{dimicsupicdakic} fully extends to quantum state certification. There a reliable certification scheme is provided for the case of independent copies to large certificates, e.g. consisting of $O(N)$ copies. Unfortunately, the full adversarial scenario is still unresolved and remains for future investigations.

 \section{Universal data records and partial tomography}
 
We have seen that in the limit of tens of copies, one can still construct powerful techniques that extract surprisingly sophisticated conclusions on unknown quantum data. We end this review by considering the limits of such extraction, that is, what is the limit of information one can know about an arbitrarily sized state given a constant number of copies of that state? Many things we might wish to know may be formulated as some kind of partial tomographic task, for example ``Is the state entangled?'' or ``Is the state $\epsilon$-close to some target quantum state?''. With so many possible questions for the same target, we might ask ourselves how many of them can be determined in parallel? Can it be done efficiently or rather, can we accurately extract multiple classical features from a moderate number of data samples?

Suppose now we introduce another ``resource'' to manage in our pursuit of efficient query protocols; our own indecision. If we do not \emph{a priori} know what classical information we wish to extract from our target quantum system, what choice of measurements maximises our knowledge at a later point? Since our choice is made \emph{a posteriori}, then all possible questions we \textit{could} ask a state are equally possible and so we must take some kind of \emph{universal data samples} that best approximates the space of all possible future queries. This is certainly achievable via full state tomography. Tomography schemes abound that aim to attack the difficulty of this task through this tactic, however a seemingly unavoidable fact of estimating properties of an arbitrary density operator is the required polynomial number of measurements in the dimension $d$. More precisely, achieving an absolute error $\epsilon$ in the estimation of an unknown density matrix requires at least ${O}(d^{2}\epsilon^{-2})$ \cite{aaronson2017} copies of a quantum state. This has to be combined with post-processing which requires storing and manipulation of exponentially large matrices. Such tasks is certainly beyond the scope for large quantum systems.

However, full state tomography may provide more information than actually needed. Our task may not require the computation of any feature of a quantum state but some more restricted class. Clearly, there is a resource-gain trade-off relation as further knowledge requires further resources, but one can get surprisingly far extracting interesting properties of the system while needing very little resources. The most significant development towards addressing this problem in recent times is due to Aaraonsons \cite{aaronson2017} breakthrough. Within it he describes a protocol dubbed ``shadow tomography'',  wherein an exponentially sized list (of mean values) of binary observables on a quantum state of dimension $d$ (mixed or otherwise) may be estimated to high precision using a measurement sample of  $O(\log d)$ size.

The name is derived from the idea that one is not especially interested in the entire quantum state but rather its projections onto a fixed set of observables- the lower dimensional ``shadows'' of a quantum state. 
With this in mind, suppose we wish to estimate a set of $M$ linear features $\{\mathrm{tr}[\rho E_1],..,\mathrm{tr}[\rho E_M]\}$ with as few copies of $\rho$ as possible. Rather surprisingly, shadow tomography shows that $M$ can be exponentially large with only a polynomial resource overhead. This statement is certainly worthy of consideration given our original problem. The main result of Aaransons paper is the following theorem:
{\bf (Aarronson \cite{aaronson2017})}
\emph{Shadow tomography is solvable using only}
\begin{equation}
    N = \tilde{O}\left(\frac{\log{1/\delta}}{\epsilon^4}\cdot \log{M}^4 \cdot \log{d}\right)
\end{equation}
\emph{copies of the target state $\rho$ where the $\tilde{O}$ hides a polylog factor. The procedure is fully explicit.}

The consequences of this should be readily apparent given the preceding review. A set of binary observables $\{E_1,...,E_M\}$ on an arbitrary quantum state can be estimated to within an $\epsilon$ absolute error with probability $1-\delta$ using a number of samples that grows logarithmically with the dimension and size of the estimated set. We direct interested readers to the original paper for proof of the above theorem and content ourselves here with answering why this does not immediately solve the problem of partial tomography. Though shadow tomography is theoretically efficient in most of the required categories, namely in terms of sample number, computational complexity and memory complexity, it unfortunately fails when considering the sophistication of the required measurements. The protocol requests joint measurements to be made on tensor products of the target state of a size $\epsilon^{-2}\log{d}$, which are repeatedly measured using carefully performed non-demolition measurements \cite{braginsky1980}, themselves a difficult procedure to perform in experiments. It is worth noting that it is not shown that these resource demands are strictly required for the protocol and indeed this was not a stated goal of the work.

\begin{figure*}[ht!]
\centering
\includegraphics[width=0.75\textwidth]{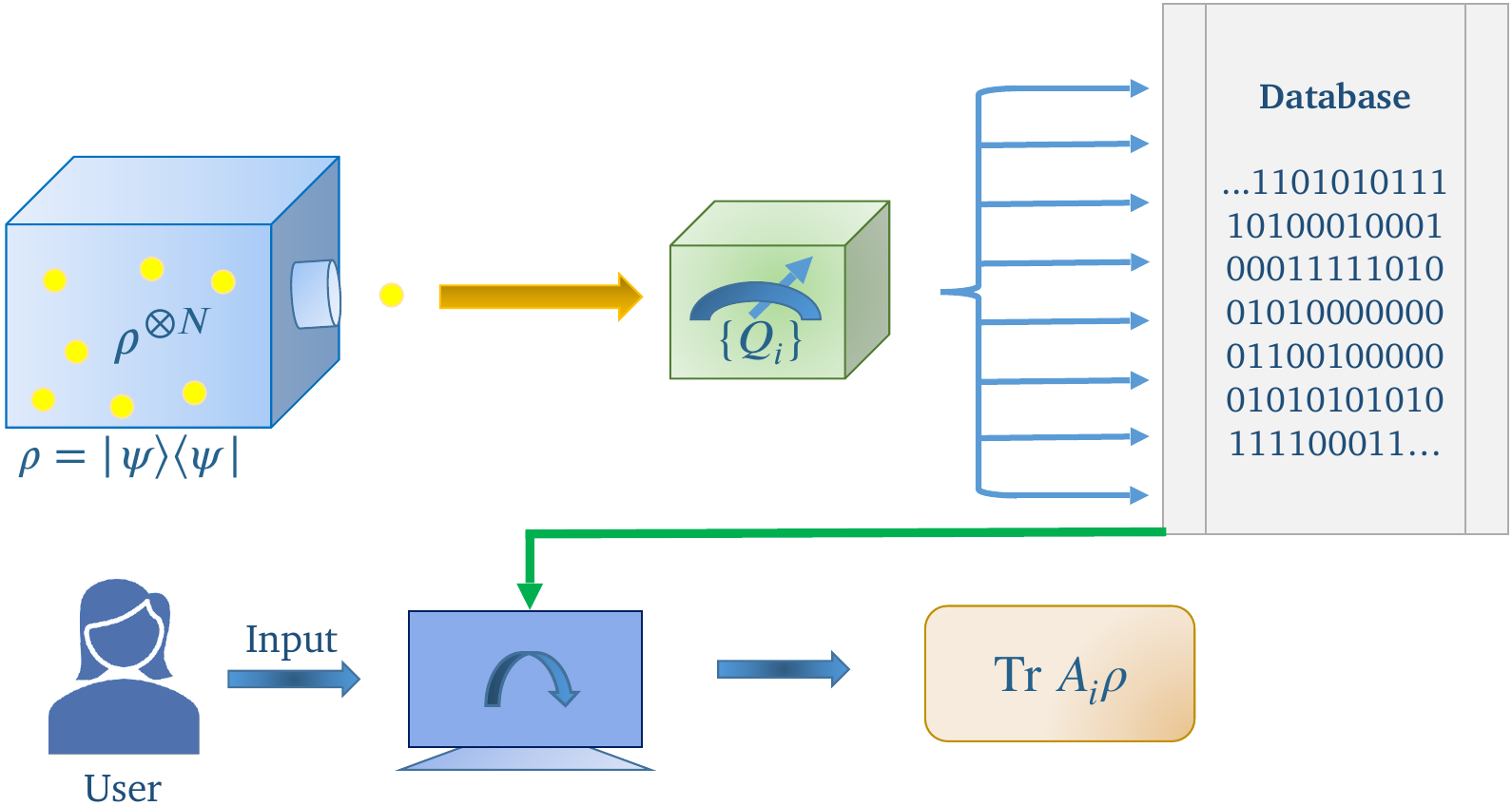}
\caption{\protect\raggedright \textbf{Protocol for estimation on demand.} The central idea is based around a two stage procedure: data acquisition and post-processing. In the first phase the universal data record of size $N$ is collected via some kind of universal POVM (e.g. information complete measurement). In the second phase, an user chooses on demand certain feature to extract (e.g. from a continuous class). A simple post-processing (low memory and computation) of the collected data furnishes the task, resulting in estimation confined to an absolute error of $O(\frac{1}{\sqrt N})$. Every new estimation (from the same data) comes at the logarithmic cost thus enabling extraction of $M$ features with the $\log M$ overhead.}
\label{binary_fig}
\end{figure*}

We review here two protocols that go beyond these limitations: \emph{selective quantum state tomography}~\cite{morris2019selective} and \emph{classical shadows}~\cite{kueng2019, huang2020predicting}.
The main emphasis here is on low-cost implementation and a universality property: we ask for the possibility of extracting \emph{on demand} (a posteriori) arbitrary features (from a given class) of a quantum state from some kind of \emph{universal data record} of moderate size. To illustrate this, suppose we wish for a protocol that allows for efficient estimation of a (finite) selection of observables from a continuous class - after our experiment is complete. On the surface this seems a monstrous request to make and one that can only begun to be fulfilled by a full state tomography. Ultimately, we shall see how this is done (for a class of bounded observables) with a cost that is completely dimension independent and requires a resource complexity that is $\log M$ for $M$ different features (a linear cost in exponentially many). The general protocol is illustrated in Figure \ref{binary_fig} and it reassembles the one defined in the introductory section with the difference being the possibility of re-using the same data (a universal record) to estimate on demand (a posteriori) a feature from some predefined (continuous) class of features. The protocol is described concretely in subsequent sections.     

\subsection{Selective quantum state tomography (SQST)}

Our task now is to weaken the stringent requirements on the measurements required for shadow tomography while still being able to estimate many operators simultaneously. To begin, let us settle for simultaneous estimation of the unit operators $A_{ij} = \outerproduct{j}{i}$, where $i=1\dots d$ and $d$ is arbitrarily large. The expectation values of these operators corresponds to the density operator element $\rho_{ij}$. A naive one-by-one measurement strategy is obviously inefficient here, as estimation of another unit operator may then require an entirely new set of measurements resulting in the general cost growing with the dimension of the system $d$. On the other hand, if one estimates various functions from the same data sample, wherein each individual estimation is efficient in the sense of a Chernoff-like bound \eqref{eff_inequality}, then we can ensure the accuracy of multiple estimations within the fixed overall error only at the logarithmic cost $\log M$ for $M$ parallel estimations (this follows from a simple union bound \cite{boole1847} for multiple random variables). This point is the crux of the protocol - once a sufficient set of measurements have been generated for a universal data record (see Figure \ref{binary_fig}), any density matrix element $\rho_{ij}$ can be estimated on demand at guaranteed precision from identical data. To do this without the complexity of measurements demanded by shadow tomography requires the introduction of a special POVM based on mutually unbiased bases \cite{durt2010}.

To construct the protocol, we shall first pick an adequate set of measurements. The set of all matrix units $A_{ij}$ forms a basis in the operator (Hilbert-Schmidt) space, thus the universal data record has to be constructed from an informationally complete POVM. The simplest and most practical choice is local measurements which are sufficient for information completeness in general but they are of limited applicability in the context of partial tomography \cite{kueng2019}. Thus one needs entangled measurements in general, keeping in mind that these shall be of a low computation complexity (i.e., implementable via low-depth quantum circuits). 

The first such basis that springs to mind is one built from mutually unbiased bases (MUB)s. MUB sets are groups of orthogonal bases defined on a finite dimensional (of dimension $d$) Hilbert space. They hold the special property whereby any two basis elements $\ket{i,m}$ and $\ket{j,n}$ drawn from different bases -- indexed as $m$ and $n$ -- have a constant inner product $|\ip{i,m}{j,n}|^2=1/d, \; \forall m\neq n$. Here $i,j=1\dots d$ index the basis elements, while $n,m=1\dots d+1$ label the basis. While there are infinitely many complete MUBs for a given dimension, we are always free to apply a global unitary to each element of the set, transforming them into a another while maintaining the inner product between elements. Due to this, we will always choose the $m=1$ basis to be the computational basis and define the remaining bases in terms of this set
\begin{equation}\label{MUB_exp}
\ket{k,m} = \frac{1}{\sqrt{d}} \sum_{l=1}^{d-1}\alpha_{l}^{km}\ket{l,1}; \quad m\neq 1,
\end{equation}
with $|\alpha_l^{km}| = 1$. The specific form of $\alpha_l^{km}$ is dependent on the dimension of the underlying Hilbert space, with different expressions for prime \cite{ivonovic1981} and prime power \cite{Paterek_opdef} dimensions. 
To proceed we use a useful fact \cite{Paterek_opdef} about arbitrary operators $A$ acting on the same space our MUB is defined upon, namely that
\begin{equation}\label{op_rep}
A = - \tr(A)\mathbb{1} + \sum_{m=1}^{d+1} \sum_{k=1}^{d} O_{k}^{(m)} \Pi_k^{(m)},
\end{equation}
with $O_{k}^{(m)} = \tr[A \cdot \Pi_k^{(m)} ]$. The $\Pi_k^{(m)}$ are constructed from the basis elements of the MUB such that $\Pi_k^{(m)} =  \outerproduct{k,m}$. The presented decomposition proofs information completeness of MUBs and we can define the corresponding POVM as$\{R_k^{(m}=\Pi_k^{(m)}/d\}$ with $k,m$ indexed as before. 

A particularly critical example may be found in the matrix unit operators. Let $A_{ij} =\outerproduct{j}{i}$ with $\ket{i}$ defined in the computational basis and $i\neq j$. Their decomposition \eqref{op_rep} adapted to the POVM elements reads
\begin{equation}
    A_{ij}=\sum_{k=1}^d\sum_{m=2}^{d+1}\eta_{ij}^{km}R_{km}.
\end{equation}
Here $\eta_{ij}^{km}=\alpha_{i}^{km*}\alpha_{j}^{km}$, thus $|\eta_{ij}^{km}|=1$ which is the crucial property. Since $\expval{A_{ij}} = \tr[\rho A_{ij}] = \rho_{ij}$, measuring a particular operator element $\rho_{ij}$ amounts to estimating the expectation value of $A_{ij}$. Given the decomposition above, the mean values $\expval{A_{ij}}$ are equivalent to the expectation value of the random variable $\eta_{ij}^{(s)}  \in \{\eta_{ij}^{km} \,|\,m= 2 \dots d+1, k= \! 1 \dots d \}$, associated with outcomes of the POVM $\{R_k^{(m)}\}$. 
Practical implementation of this POVM amounts to randomly choosing one of $d$ orthonormal basis sets (not including $m=1$) to measure a copy of $\rho$ in, each with probability $1/d$ of being selected. A tomography to estimate $\rho_{ij}$ would then proceed by the generation of $N$ copies of $\rho$, each measured using this POVM. For each measurement outcome, indexed by $s$, we update an approximation to the above sum as the following estimator
\begin{equation}\label{el_est}
    \rho_{ij}' = \frac{1}{N}\sum_{s=1}^N \eta_{ij}^{(s)}.
\end{equation}
To be completely explicit, a selective quantum state tomography would proceed in experiment as follows: 
\begin{enumerate}
    \item{Measure a copy of the quantum state $\rho$ using the POVM defined by $\{R_k^{(m)} \}$, to get the measurement result $(k,m)$.}
    \item{Repeat the procedure $N$ times to get the (universal) measurement record of outcomes $\{(k_1,m_1),\dots (k_N,m_N) \}$. This concludes the experimental phase of the SQST.}
    \item{In post processing, chose a particular element $\rho_{ij}$ to compute $\eta_{ij}^{(s)}$ and the sum in Eq. \eqref{el_est}.}
    \item{To estimate a different element $\rho_{st}$, simply update the values of $i,j$ to $s,t$ and recompute the estimator, without further measurements.}
\end{enumerate}

If we calculate the number of state copies $N$ of $\rho$ required for the estimator $\rho_{ij}'$ to converge to $\rho_{ij}$ within some error $\epsilon$ and failure probability $\delta$. Though $\eta_{ij}^{(s)}$ is complex, we may still apply the usual concentration inequalities by considering $\eta_{ij}^{(s)}$ as two bounded random variables such that $|\Re[\eta_{ij}^{(s)}]+i\Im[\eta_{ij}^{(s)}]| = 1$ . Recall that $\rho_{ij}' = N^{-1}\sum_{s=1}^N \eta_{ij}^{(s)}$ and note that  $\mathbb{E}[\rho_{ij}'] = \rho_{ij}$. Following a concentration inequality approach we wish to compute the bound $ \Pr(\left| \rho_{ij}' - \mathbb{E}[\rho_{ij}'] \right|\geq \epsilon)$. First, we will isolate the real and complex components of the random variable $\eta_{ij}^{(s)}$. By the triangle inequality we have that
\begin{equation}
\Pr(\left| \rho_{ij}' - \mathbb{E}[\rho_{ij}'] \right|\geq \epsilon) \leq \Pr(|A|\geq \frac{\epsilon}{\sqrt{2}}) + \Pr(|B|\geq \frac{\epsilon}{\sqrt{2}}),
\end{equation}
for $A = \Re(\rho_{ij}') - \mathbb{E}[\Re(\rho_{ij}')]$ and $B = \Im(\rho_{ij}') - \mathbb{E}[\Im(\rho_{ij}')]$. 
From here, we may apply a standard Hoeffding's inequality for bounded random variables to each term individually to get
\begin{equation}\label{eq:POVM_bnd}
    \Pr(\left| \rho_{ij}' - \mathbb{E}[\rho_{ij}'] \right|\geq \epsilon) \leq 4e^{-\frac{N\epsilon^2}{2}} = \delta.
\end{equation}
We may then deduce the number of copies $N = {O}(\epsilon^{-2} \log \delta^{-1})$ required to estimate $\rho_{ij}$ with an error bound $|\rho_{ij}' - \rho_{ij}| < \epsilon$ that occurs with probability greater than $1-\delta$. This is in tandem with an ${O}(N)$ complexity overhead in both the required memory and computation, given we need only to store the outcomes of each measurement and the summation may be computed piece-wise.
For estimation of any $\rho_{ij}$, we need also to account for the diagonal case $i = j$, something we neglect in the above formulation of SQST. Fortunately the estimation of the diagonal elements of $\rho_{ii}$ is straightforward. This stems from the fact that diagonal estimation of density operators is something of a simple case, achievable with measurement in the computational basis. For truly arbitrary estimation of the elements of a density operator we thus need to maintain two
measurement records; one for the diagonal elements which gives the $\rho_{ii}$ directly, and another for the off diagonals $\rho_{ij}$, both requiring $N = {O}(\epsilon^{-2} \log \delta^{-1})$ copies of the state. Finally, an additional factor must be included if multiple elements are $\rho_{ij}$ to be estimated, corresponding to $M$ repetitions of step 4 in the experiment. This amounts to $\log M$ overhead which comes from the union bound resulting in $N = {O}(\epsilon^{-2} \log \delta^{-1}\log M)$ repetition. Remarkably, this scaling is free of the dimension $d$.

\subsection{Relation to full tomography and arbitrary observables}

It is tempting to conclude that if one case efficiently estimate all individual elements of a density operator efficiently then one can estimate the density operator itself efficiently. This is true but only in a technical sense - while SQST will give a bounded error on individual elements with high probability, the overall error of the estimated quantum state in the usual metrics - namely trace distance - may be exponentially large. This comes from SQST estimation error being equivalent to the max norm $||E||_{\max}\coloneqq \max_{ij} |E_{ij}|\leq \epsilon$ which is related related to the trace distance norm via
\begin{equation}\label{eq:bounds}
    \frac{1}{\sqrt{d^3}}||E||_1 \leq ||E||_{\max} \leq ||E||_1.
\end{equation}
This is rather unsurprising as anything else would imply a protocol that outperforms provably optimal full state tomography \cite{aaronson2017}. Of course, it is still possible to perform state tomography in the supremum norm. In a similar manner to maximum likelihood estimation, a semidefinite program 
\begin{equation}\label{eq::opt_est}
\rho_p \coloneqq \argmin_{\sigma \succeq 0} ||\rho_{L} - \sigma||_{\max},
\end{equation}
may be constructed that yields positive semi definite solutions from the data record generated by SQST \cite{morris2019selective}. Though running such an optimisation program would not be computationally efficient, the required sample complexity for all $d^2$ elements remains efficient at $\log d^2=2\log d$. 

Another interesting point to investigate is the application of SQST to estimate mean values of observables going beyond matrix units $\outerproduct{j}{i}$. 
 
Consider a general decomposition given in Eq. \eqref{op_rep} of an operator $A$
\begin{equation}
    A = \sum_{k=1}^d a_{k1}\Pi_{k}^{(1)}+\sum_{m=2}^{d+1}\sum_{k=1}^d a_{km}\Pi_{k}^{(m)}=A_0+\tilde{A}, 
 \end{equation}
where we intentionally separate decomposition into computational basis which gives diagonal matrix $A_0$ and the rest of $\tilde{A}$ with all 0 on the main diagonal. Furthermore, we restrict our attention to operators bounded in entrywise 1-norm $||A||_1=\sum_{ij}|a_{ij}|$, where $a_{ij}$ are matrix elements of $A$ in the computational basis. Given $||A||_1$ bounded we have all elements $|a_{ij}|\leq ||A||_1$ also bounded. As before, the estimation is broken into two stages: estimation of $A_0$ which is efficiently done in computational basis (since $a_{ii}$ are bounded) and estimation of $\tilde{A}$ which is performed by random sampling of MUBs (see previous section). The corresponding random variable $a_{km}$ is bounded, i.e., $|a_{km}|=|d\tr[A'\Pi_{k}^{(m)}]|\leq d\sum_{ij}|a_{ij}||\ip{i,1}{k,m}\ip{k,m}{j,1}|\leq ||A||_1$, thus the efficiency of the estimation follows from the Hoeffding bound of Eq. \eqref{eq:POVM_bnd} with $N = {O}(\epsilon^{-2} \log \delta^{-1}||A||_1^2)$.

The previous analysis shows that operators bounded in entry-wise $l_1$-norm can be efficiently estimated by the SQST procedure. However, these bounds are not optimal. To see this, suppose we simultaneously estimate the mean values of $4^n-1$ Pauli operators (excluding identity) $A=\sigma_1\otimes...\otimes \sigma_n$, where $\sigma_k$ is one of the standard Pauli matrices. We have $||A||_1=d=2^n$, thus our previous analysis predicts a sample cost of $N={O}(4^n n)$, where the factor $n\sim\log 4^n$ comes from the union bound. However, it is well known \cite{anton2002} that the set of $4^n-1$ Pauli operators can be factored into $2^n+1$ groups each composed of $2^n-1$ commuting operators with their common eigenbases being MUBs. This means that a single MUB measurement can return all $2^n$ mean values (of commutative Paulis) at the cost of ${O}(n)$ thus the estimation of all $4^n$ requires ${O}(2^n n)$ copies (to measure all in MUBs). This scaling is known to be optimal \cite{crawford2021}. Consequently, this is quadratically better than the estimation given by the norm $||\cdot||_1$ analysis meaning that the derived bounds can be further improved. One way of doing this is to employ the Bernstein's inequality \cite{bernstein1946,zanten2001} which controls also the variance of the random variable thus leading to potentially better bounds. Another possibility to generically improve the scaling is to change the POVMs and type of estimator, e.g. instead of a simple linear estimator, one may use the median of means estimator \cite{lerasle2019}. This coincides with the next and final scheme in terms of sample complexity and is superior in terms of measurement complexity and efficiency for estimation of a general observable bounded in Frobenius norm. Along with SQST the next scheme called classical shadows \cite{kueng2019,huang2020predicting} is an entirely new regime of partial tomography not previously possible.
%\textcolor{red}{[[Ref both Richard's papers]]} 

\subsection{Classical shadows}
With shadow tomography suggesting the possibility of a sample-efficient universal algorithm and SQST demonstrating that a degree of generality can still be achieved with vastly simpler measurements, we close this review with the current state of the art in efficient quantum tomography. Considering again the protocol above, we defined an alternative scheme using a generalised measurement basis - the mutually unbiased bases, producing a partial tomography protocol that can construct many independent linear functions on a target state while remaining resource-efficient.

One now wonders why this was the case - a choice of unbiased bases as a first target for universal measurements is intuitive given that they form an informationally complete POVM and their very nature of containing minimal measurement bias, but they work unexpectedly well for an educated guess. A possible reason for this lies in a so-far unmentioned MUB property, namely that they form a $t$-design of degree two \cite{rotteler2005mubtdesign}. While a full description of $t$-designs is unnecessary here (see Ref. \cite{eisert2007tdesign} for a complete treatment in the context of quantum mechanics), it is sufficient to understand that a quantum $t$-design is a probability distribution that approximates polynomial functions of order $t$ over the complete distribution for some set. A simple (classical) example are the average of some polynomial function over the real sphere.

The relevance of this here is that such designs can be used to approximate the probability distributions of a generalised measurement basis. Higher order designs better reproduce the key properties of a distribution with a two design correctly producing the same expectation value and a three design correctly showing the same sample variance. The natural and immediate question is what do higher order $t$-designs yield? We clearly see from the Bernstein inequality that the variance of an observable plays a heavy role in terms of the efficiency of an estimator, so one may presume that a $t$-design that reproduces both the correct expectation value and variance of the approximated distribution will have improved performance again.

Coupled with a statistical trick known as the median-of-means \cite{lerasle2019}, this is the strategy of Keung et al. \cite{kueng2019} who show that through randomised Clifford measurements (a three-design) they are able to estimate $M$ observables at a number of samples that grows as 

\begin{equation}
    N=O\left(\frac{||A||_{max}^2\log M}{\epsilon^2} \log\delta^{-1}\right)
\end{equation}
with $||A||_{max}=\max(||A_1||_2\dots,||A_L||_2)$ being the maximum two-norm (Frobenius) of the $M$ observables to be estimated. Included within this bound are entanglement witnesses and fidelity estimation, both of which can be performed efficiently regardless of the system size. With regards to a two design, a three design (when coupled with sufficient statistical methods) is slightly more expensive in terms of gate complexity, requiring a cubic number of Clifford gates to achieve sufficient randomness over the Haar measure as compared to the linear cost of generating MUB measurements. Both may be considered computationally efficient however and one gains a powerful advantage when the use of a three-design Clifford measurement is allowed.

 \section{Concluding remarks}
 
In this work, we have reviewed recent approaches to answering queries of quantum states of increasing size, while avoiding an unacceptable overhead in resources. By first considering efficient tomography to be a series of queries that become exponentially unlikely to pass for all states excepting those that answer positively, we showed how this leads to hyper-efficient protocols. We demonstrated this through high-performance entanglement detection using a single copy of a quantum state; a counter-intuitive result for an estimation protocol. This was then extended by showing how the same protocol can be used for cluster states, a specific class of quantum state and the ground states of local Hamiltonian. 

We proceeded to the case where a limited number of state copies is available, one can work in the few-copy regime and observe the presence of entanglement in the state with a protocol to translate any entanglement witness into a probabilistic framework. We showed that this scenario is well-suited for experimental implementations by reviewing an application to a photonic six-qubit cluster state. By demonstrating that the method provides the ability to detect quantum entanglement with very high confidence with only about hundreds of state copies, the extremely low requirements in terms of time and experimental resources were confirmed.

With experimental viability in mind, we gave a description of shadow tomography which set the stage for Selective Quantum State Tomography, showing how a special choice of POVM leads to the efficient estimation of a wide class of linear quantum functionals. This in turn leads to the current state of the art for partial tomography, a $t$-design based protocol using the classical shadows of a quantum state which leads to efficient estimation of an exceptionally large class of observables.  

This high performance is most clearly seen in the context of possible partial tomographies performed; namely fidelity estimation (where the observable is another density operator), entanglement witnesses and entropies, correlation functions up to order two and the energies of many-body local Hamiltonians.

Beyond the methods presented in this review, it is fair and also worth mentioning novel techniques that instead employ machine learning to reduce the verification requirements. In fact, the use of machine learning for quantum applications is in general experiencing rapid progress and proving useful in tasks like entanglement detection using neural networks \cite{roik2021accuracy,yosefpor2020finding} or unsupervised learning \cite{chen2021detecting}, and quantum state tomography using neural networks \cite{torlai2018neural}. It is also relevant that a comparable method (to SQST) for estimating elements of a density matrix exists in the continuous variable (CV) regime. Here, it is known that the estimation error depends directly on the energies\! \cite{dariano1996,gheorg2018}, i.e., the estimation error for a matrix element $\rho_{nm} $ increases with $n$ and $m$ ($n,m$ index the energy eigenstates). Notably, the same behaviour is not observed in SQST of discrete systems which forms a point of interest for developing tomographic strategies targeting CV systems.

Our main focus in this review was on sampling (in terms of measurement complexity) where the presented techniques exhibit a dimensional independence, a property that is crucial for real application. There are however a number of open questions that remain to be addressed in future work. In the context of entanglement detection one immediately realises that verification models tend to be tailored to detect entanglement in the vicinity of a target state which requires some prior knowledge of the state preparation. Which witnesses and corresponding verification procedure should one use then if there is no such prior knowledge? This is an open research topic and not many results may be found in the literature, owing to the difficult nature of this restriction. In such cases, one promising direction may be to use the method of so-called random correlations~\cite{tran2015, tran2016, ketterer2019}, which was developed for entanglement detection and try to incorporate it into the decision-theoretic framework presented here.

Another pressing issue is the assumption of ``IIDness'' (identical and independently distributed) samples which is highly questionable in  the context of near term quantum devices given high error rates, source drifts and lack of control and manipulation. Our entanglement detection schemes surpass the IID limitation by employing random sampling techniques, but difficulties arise immediately at the next level of sophistication \'{a} la quantum state verification. One can mitigate this issue via conditional fidelities \cite{dimicsupicdakic, JD}, but it remains an open question whether some nontrivial statements can be made about the full state produced by the source. A possible way out may be found in the de-Finetti reduction theorems \cite{christandl2012reliable}, or with the help of entropy accumulation theorems \cite{dupuis2016entropy,Arnon_Friedman_2019} where resorting to permutational invariance is not allowed. Another option that may follow form our single-copy framework is to fold all accessible ((non-IID) copies setting into a large single-copy and perform verification in a single-copy scenario. While this seems to be reasonable option, what remains to be clarified is: {\it what is the class of states and properties that admit reliable single-copy verification/estimation?} Our protocols reviewed here are the first steps towards answering this question. In this way, there is another conceptual issue to be addressed that concerns the operational meaning of physical quantities in a single-shot scenario.

A particularly pertinent open question, especially in the context of near term quantum devices is the trade-off between measurement complexity and the corresponding increase or decrease of efficiently estimable quantities. As noted in Refs. \cite{morris2019selective, kueng2019}, the power of these techniques appears to be uniquely sourced from the choice of measurements performed. Specifically that they are two (in SQST) and three (in tomography) designs in $t$-design parlance \cite{eisert2007}. In particular, when estimators in classical shadow tomography are constructed from local measurements only, i.e., a one-design, the performance of the scheme drops significantly.  Such a question was considered in the original work of classical shadows \cite{kueng2019} in the context of Pauli measurements, finding the complexity scaled unsurprisingly in the non-locality of the target observable. It also is something of the worst case scenario in that one is restricted to a fixed set of weak measurements. Instead one may introduce adaptability into the POVM implemented in the measurement phase of a scheme as was done by Garc\'{i}a-P\'{e}rez et al. \cite{guillermo2021}. Despite the optimisation introducing increased classical post-processing into the protocol, it does not compromise the circuit complexity of the POVM. It remains less powerful than a complete shadow tomography but demonstrates high performance on the limited but highly relevant class of variational quantum eigensolver (VQE) problems \cite{peruzzo2014variational}. 

This is a well chosen compromise, since Clifford and MUB measurements are not trivial to implement owing to the inclusion of control operations between arbitrary subsystems, it is highly desirable to find similar reductions with perhaps different compromises being found for different problem instances. While certainly worth pursuing, this can be seen as equivalent to constructing POVMs that approximate a $t$-design of some order using a simpler set of generators that the Clifford group. Given that finding $t$-designs in the first place is already difficult, this is a challenging task. 

With all these questions in mind, it appears that the time is nigh for an exciting new class of tomographic protocols, ones without the apparent drawbacks that have plagued state tomography since its inception allowing for direct probing of quantum systems in the NISQ technology regime and beyond.

\medskip
\textbf{Acknowledgements} \par %delete if not applicable))
J.M. and B.D. acknowledge support from the Austrian Science Fund (FWF) through BeyondC-F7112. V.S. acknowledges support from the FWF through BeyondC-F7113. A.G. acknowledges funding provided by the Faculty of Physics, University of Belgrade, through the grant by the Ministry of Education, Science and Technological Development of the Republic of Serbia. 
 
\medskip
\textbf{Conflict of Interest} \par
The authors declare they have no conflict of interest.

\bibliographystyle{unsrturl}
\bibliography{biblio_MSP}
\end{document}